\documentclass[reprint,aps,prb,nofootinbib,superscriptaddress,longbibliography]{revtex4-2}
\usepackage{amsmath}
\usepackage{amssymb}
\usepackage{graphicx}
\usepackage{dcolumn}
\usepackage{bm}
\usepackage{comment}
\usepackage{color}
\usepackage{hyperref}
\usepackage{xr-hyper}
\usepackage[autostyle, english = american]{csquotes}
\MakeOuterQuote{"}

\newcommand{\vect}[1]{\boldsymbol{#1}} 

\begin{document}

\title{Electron localization, charge redistribution, and emergence of topological states at graphite junctions from Green's function embedding}

\author{L. Soneji}
\affiliation{Department of Physics, University of Bath, Claverton Down, Bath, BA2 7AY, United Kingdom}
\author{S. Crampin}
\affiliation{Department of Physics, University of Bath, Claverton Down, Bath, BA2 7AY, United Kingdom}
\author{M. Mucha-Kruczyński}
 \email{M.Mucha-Kruczynski@bath.ac.uk}
\affiliation{Department of Physics, University of Bath, Claverton Down, Bath, BA2 7AY, United Kingdom}

\date{\today}

\begin{abstract}
Low-energy electronic behavior in graphite crystals is highly dependent on the relative stacking arrangement of the constituent layers. Topologically non-trivial electronic states can arise due to interrupted rhombohedral (ABC) stacking, localized at the edges of the stacking region, but not in the case of Bernal (AB) stacking. Here, we study the electronic properties of junctions between half-crystals of graphite of either Bernal or rhombohedral stacking, using a charge self-consistent tight-binding method and embedding potentials to account for the influence of layers far from the junction. We find junction-localized electronic states to be a ubiquitous feature, and all systems but one involving a rhombohedral half-crystal support a flat-band expected to exhibit electronic instabilities and strongly-correlated states. Nascent flat-band states associated with finite rhombohedral stacking sequences extend the physics into pure Bernal systems.
\end{abstract}

\maketitle

\section{\label{sec:introduction}Introduction}

Graphite, the bulk form of stacked carbon layers bonded by weak van der Waals (vdW) forces, has been of scientific interest for over 60 years \cite{McClure1957, Slonczewski1958}. This interest has partially been due to the remarkable electronic behavior observed near the Fermi level, originating from the stacking of the layers \cite{Charlier1991}. The two most common structural types of graphite are Bernal (AB) and rhombohedral (ABC) stacked: infinitely periodic crystals with a unit cell dictated by the position of the third layer. Bernal graphite is considered to be the more energetically stable of the two, with the vast majority ($\sim80\%$) of all naturally occurring graphite found in the Bernal form, and $\sim15\%$ in the rhombohedral form \cite{Lipson1942, Haering1958, Lui2011}. The former can be transformed into the latter via mechanical shear \cite{Laves1956}, and the small formation energy difference $\sim0.1$ meV/atom between the two \cite{Savini2011, Nery2021} results in significant ($\sim5\%$) stacking disorder in natural crystals. Following the exfoliation of graphene and recognition of its remarkable properties \cite{Novoselov2004}, developments in the fabrication and processing of graphite materials have developed considerably, enabling the synthesis and study of high-quality crystals with thicknesses from one to hundreds of layers \cite{Novoselov2004, Mak2010, Lui2011, Yin2019, Shi2020}, as well as control and manipulation of their stacking order \cite{Yang2019, Pan2021, Wu2021, Yeo2025} and rotational alignment \cite{Hu2022}.

Motivated by these capabilities, we study here the emergent low-energy electronic properties of junctions formed between two lattice-matched graphite crystals, allowing each to be either Bernal or rhombohedral stacked. Several previous studies of such systems exist \cite{Min2008, Arovas2008, Taut2013, Koshino2013, Taut2014, Taut2016, GarciaRuiz2019, Shi2020, Muten2021, GarciaRuiz2023, Sarsfield2025, Soneji2026}, especially of stacking faults arising when both half-crystals have the same structure. These typically consider one or two junction structures, modeled as periodically-repeated or in thin-film geometries. Here, we consider a comprehensive set of 12 junction configurations, including those involving AA stacking at the interface --- although energetically unfavored, local AA-stacking naturally arises in twisted low-angle system \cite{Cisternas2012, Dey2023} so our results provide a reference point for interpreting the properties of such systems. Using a Green's function-based tight-binding approach in which the component crystals are semi-infinite, we extract the intrinsic properties of the junctions, isolated from nearby boundaries and junction-junction interactions. This allows us to identify a wealth of junction-localized states: all non-trivial junctions possess such states, and all but one host a junction state contributing spectral features near the Fermi energy. Using a simplified version of the tight-binding theory we supplement our numerical results with analytic analysis of dispersion relations and of the spatial distribution of flat band states. The latter, associated with rhombohedral half-crystals, point to a high likelihood of electronic instabilities and strong correlation effects at junctions, with nascent flat-band states associated with finite rhombohedral stacking sequences extending the physics into pure Bernal systems.

The outline of our work is as follows: In Section \ref{sec:graphite} we describe the atomic structure of the graphite junctions studied, and in Section  \ref{sec:model} present the electronic model adopted. Section \ref{sec:greensfunctions} describes the Green's function method used to obtain the junction electronic structure and Section \ref{sec:charge} the approach that we take to deal with charge redistribution. Section \ref{sec:ES} summarizes key results of the electronic structure, and we conclude with a discussion in Section \ref{sec:discussion}.

\section{\label{sec:graphite}Graphite junctions}

\begin{figure}[t]
\centering
\includegraphics[width=0.95\columnwidth]{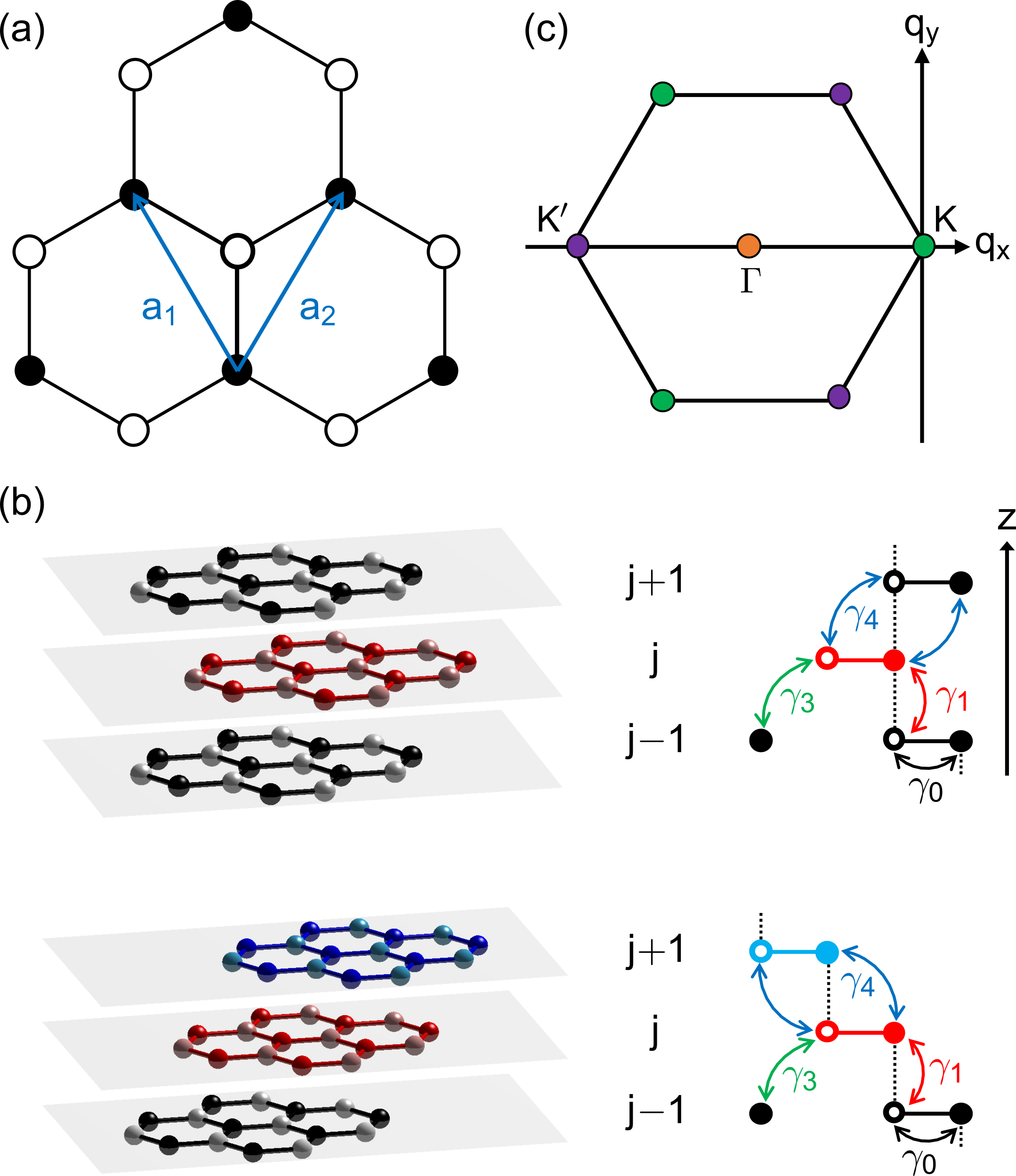}
\caption{\label{fig:structure} 
a) In-plane real space structure of a graphene monolayer. Filled (empty) circles indicate the positions of atoms belonging to sublattice $A$ ($B$). b) Trilayer sections of Bernal (upper) and rhombohedral (lower) stacked graphite. Hopping processes included in the Hamiltonian used to model the graphite are denoted by $\gamma_n$, where $n=0,1,3$, and $4$. Here the out-of-plane $z$-axis runs from bottom to top. c) The hexagonal first Brillouin zone of graphene, side length $4\pi/(3a)$. Equivalent high-symmetry points on the zone edges are denoted with like colors.}
\end{figure}

Graphite consists of stacked layers of graphene: a honeycomb lattice of carbon atoms occupying two non-equivalent sublattices labeled $A$ and $B$, as shown in Fig.\@ \ref{fig:structure}a. The unit cell of Bernal graphite contains two graphene layers positioned such that an $A$ atom lies directly above a $B$ atom --- see top row of Fig.\@ \ref{fig:structure}b. Exactly half of the atoms in the crystal are directly aligned in the out-of-plane $z$ direction, and form the "backbone" of the crystal; the remaining half instead possess dangling bonds above and below. The structure is described by primitive lattice vectors $\vect{a}_1 = a(-1/2,\sqrt{3}/2,0)$, $\vect{a}_2 = a(1/2,\sqrt{3}/2,0)$, and $\vect{c}_{\textrm{AB}} = c_{0}(0,0,2)$, where $a = 2.46$ \AA\ is the in-plane lattice constant and $c_{0} = 3.35$ \AA\ is the interlayer distance \cite{Dresselhaus1981}. In rhombohedral graphite, the layer placement is repeated for the third and subsequent layers --- see bottom row of Fig.\@ \ref{fig:structure}b, giving a three-layer unit cell with $\vect{c}_{\textrm{ABC}} = c_{0}(0,0,3)$, or one-layer primitive cell with $\vect{c}_{\textrm{ABC}} = (0,a/\sqrt{3},c_{0})$. Each atom is aligned with one neighbor in one adjacent layer so that all atoms experience the same electronic environment in the bulk.

\begin{figure*}[t]
\centering
\includegraphics[width=0.95\textwidth]{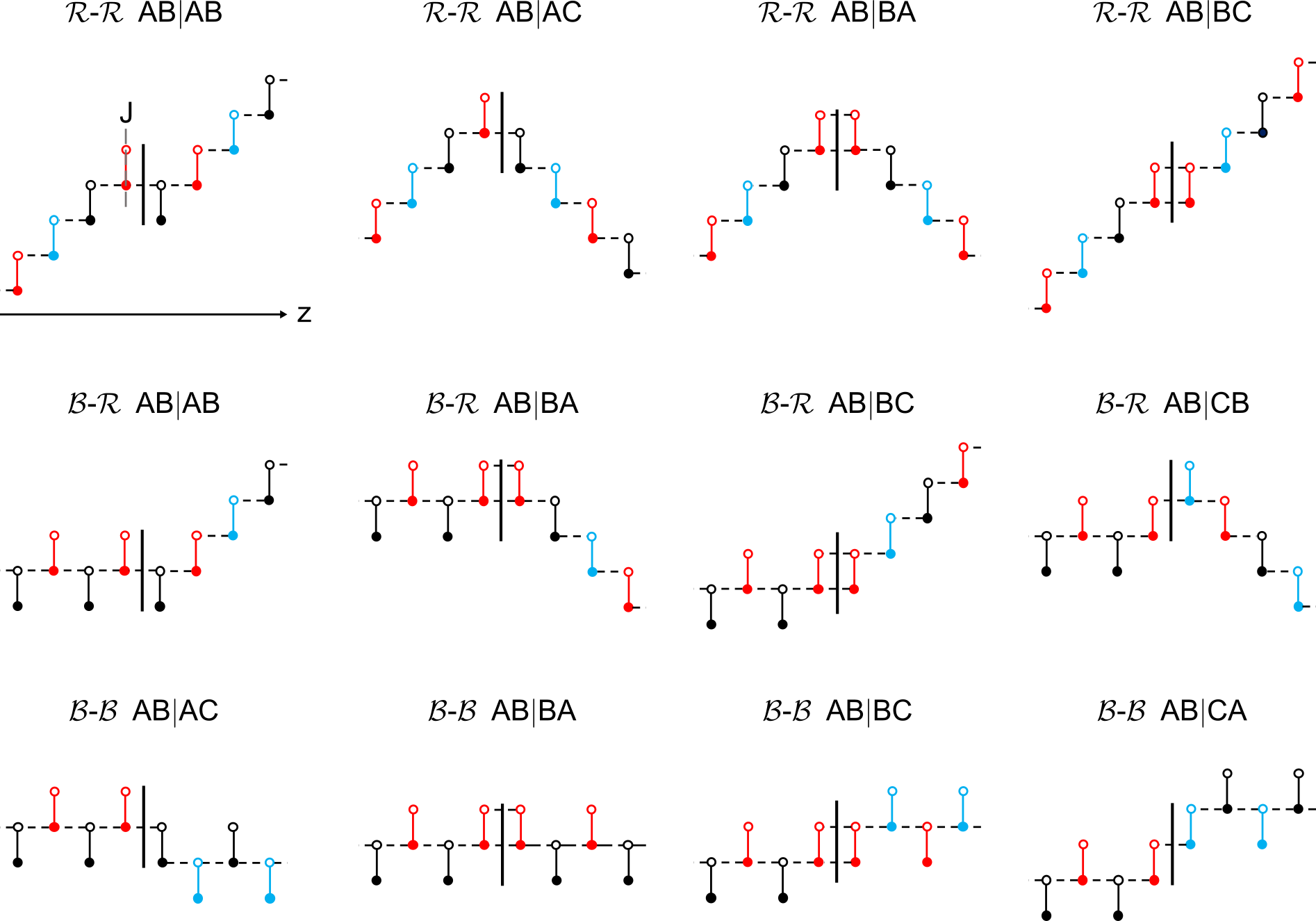}
\caption{\label{fig:crystals}
The interfacial region of the 12 distinct crystals formed by commensurate alignment of the surfaces of two half-crystals of rhombohedral ($\mathcal{R}$) or Bernal ($\mathcal{B}$) graphite. We refer to the layer directly to the left of the physical interface (the right-most layer of the left half-crystal) as layer $J$ -- see top-left panel. The out-of-plane $z$-axis runs from left to right.}
\end{figure*}

We consider the crystals formed from fully commensurately-aligned graphite half-crystals of either rhombohedral ($\mathcal{R}$) or Bernal ($\mathcal{B}$) stacking. In what follows, we consider the $z$-axis to point from left to right. Without loss of generality we let the left half-crystal terminate at the junction with AB stacked layers. Restricting the layer stacking between half-crystals to either rhombohedral, Bernal, or simple-hexagonal (AA-type stacking), there are then six possibilities for the stacking sequence of the first two layers of the right half-crystal (AB, AC, BA, BC, CA, and CB), which can be continued appropriately depending on the bulk stacking of the right half-crystal. In total, the three distinct half-crystal combinations $\mathcal{R}$-$\mathcal{R}$, $\mathcal{B}$-$\mathcal{R}$, and $\mathcal{B}$-$\mathcal{B}$, along with the six possible stacking sequences, result in 18 junctions. Two of these correspond to perfect rhombohedral and Bernal bulk graphites: $\mathcal{R}$-$\mathcal{R}$ AB$|$CA and $\mathcal{B}$-$\mathcal{B}$ AB$|$AB, respectively. By symmetry, the AB$|$AC and AB$|$CB junctions are equivalent when the constituent half-crystals are of the same stacking type. Finally, for $\mathcal{B}$-$\mathcal{R}$ stacking, the AB$|$AB, AB$|$AC, and AB$|$CA junctions are equivalent, corresponding to identical stackings translated in the stacking direction by either one or two layers. Overall, there are 12 distinct aperiodic systems, illustrated in Fig. \ref{fig:crystals}.

\section{\label{sec:model}Electronic Model}

Translational periodicity is preserved in the plane of the graphene layers, meaning the electronic states possess in-plane wave vector $\vect{q}=(q_{x},q_{y})$ as a good quantum number. We employ a tight-binding framework \cite{Dresselhaus1981, Partoens2006} in the basis $\{ \dots, \left|\Psi^{A}_{j-1}\right>, \left|\Psi^{B}_{j-1}\right>, \left|\Psi^{A}_{j}\right>, \left|\Psi^{B}_{j}\right>, \left|\Psi^{A}_{j+1}\right>, \left|\Psi^{B}_{j+1}\right>, \dots \}$ to model our system, where $\left|\Psi^{\mu}_{j}\right>$ is the sublattice Bloch state formed from $p_z$ orbitals on site $\mu$ of layer $j$. The resulting Hamiltonian, suitable for describing the low-energy electronic properties, takes the form
\begin{equation} \label{eqn:hamiltonian}
H = 
\begin{pmatrix}
\ddots & & & & \\
	& \hat{H}^{0}_{j-1} & \hat{V}_{j-1,j} & 0 & \\
	& \hat{V}_{j-1,j}^{\dagger} & \hat{H}^{0}_{j} & \hat{V}_{j,j+1} & \\
	& 0 & \hat{V}_{j,j+1}^{\dagger} & \hat{H}^{0}_{j+1} & \\
 & & & & \ddots
\end{pmatrix}.
\end{equation}
We use a hat to denote the $2 \times 2$ matrix blocks comprising the Hamiltonian; the diagonal blocks are graphene monolayer Hamiltonians with stacking-dependent on-site energy shifts $\Delta^{\mu}_{j}$,
\begin{equation} \label{eqn:monolayer}
\hat{H}_{j}^{0} = 
\begin{pmatrix}
\Delta^{A}_{j} & -\gamma_{0} f_{\vect{q}}\\
-\gamma_{0}f_{\vect{q}}^{*} & \Delta^{B}_{j}
\end{pmatrix},
\end{equation}
where $f_{\vect{q}} = 2 \exp{\left(-i\tfrac{q_{y}a}{2 \sqrt{3}} \right)} \cos{\left( \tfrac{q_{x}a}{2} + \frac{2\pi}{3} \right)} + \exp{\left(i\tfrac{q_{y}a}{\sqrt{3}} \right)}$.
The off-diagonal blocks in Eq.\@ (\ref{eqn:hamiltonian}) describe interlayer coupling. When two adjacent layers are AB (or BC or CA) stacked with respect to each other, this is
\begin{equation} \label{eqn:pertubation}
\hat{V}_{j,j+1} = \hat{V}_{\textrm{AB}} =
\begin{pmatrix}
\gamma_{4} f_{\vect{q}} & \gamma_{3} f_{\vect{q}}^{*}\\
\gamma_{1} & \gamma_{4} f_{\vect{q}}
\end{pmatrix}.
\end{equation}
When they are BA (or CB or AC) stacked, $\hat{V}_{j,j+1} = \hat{V}_{\textrm{BA}} = \hat{V}_{\textrm{AB}}^{\dagger}$. When two adjacent layers are perfectly aligned with one another (AA or BB or CC stacking),
\begin{equation} 
\hat{V}_{j,j+1} = \hat{V}_{\textrm{AA}} =
\begin{pmatrix}
\gamma_{1} & \gamma_{3} f_{\vect{q}} \\
\gamma_{3} f_{\vect{q}}^{*} & \gamma_{1}
\end{pmatrix}.
\end{equation}
We use $\gamma_{0}=3.16$\ eV, $\gamma_{1}=0.38$\ eV, $\gamma_{3}=0.29$\ eV, and $\gamma_{4}=0.13$\ eV based upon a number of sources considering both $\mathcal{B}$ and $\mathcal{R}$ stacking types \cite{Dresselhaus1981, Chung2002, Shi2020, Slizovskiy2019, Cea2022, Kaladzhyan2021}. For layers distant from the junction where the Hamiltonian becomes that of the bulk, we use $\Delta^{\mu}_{j}=-4.57$\ meV for Bernal and $\Delta^{\mu}_{j}=-32.3$\ meV for rhombohedral stacked layers, ensuring aligned Fermi energies $E_{F} = 0$. Values of $\Delta^{\mu}_{j}$ differ in the vicinity of the junction due to charge redistribution --- see Section \ref{sec:charge}.

\section{\label{sec:greensfunctions} Interface Green's Function}

\begin{figure}[t]
\centering
\includegraphics[width=0.95\columnwidth]{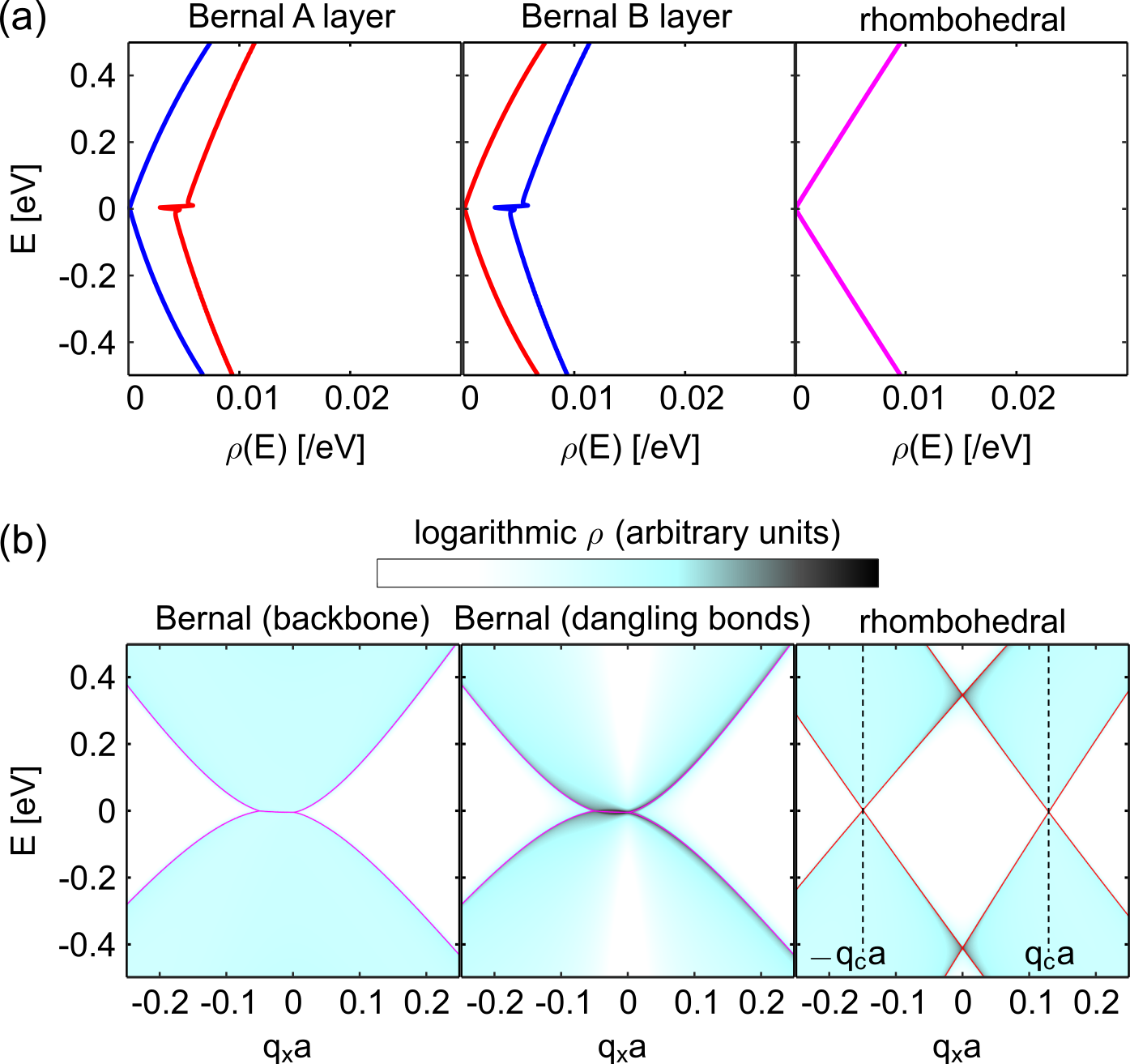}
\caption{\label{fig:bulkdos} a) The site-resolved density of states of bulk graphite of both stackings. Results for sites belonging to sublattice $A$ ($B$) are shown in red (blue), and magenta when they are equal. b) The $\vect{q}$-resolved density of states of bulk graphite of both stackings. In these results, $q_{y}=0$. Darker areas correspond to higher LDOS, and the Bernal (rhombohedral) bulk continuum is bordered by a thin magenta (red) line. Unless otherwise stated, all LDOS results presented in this document use the same scale and color scheme as used in this figure.}
\end{figure}

The Hamiltonian in Eq.\@ (\ref{eqn:hamiltonian}) describes a system of layers extending to infinity on either side of the junction, but our focus is on the electronic structure in its immediate vicinity. We calculate this via the Green's function, formally \cite{Economou}
\begin{equation} \label{eqn:Gdefinition}
G(\omega,\vect{q}) = \left( \omega - H(\vect{q}) \right)^{-1},
\end{equation}
where $\omega$ is the complex energy parameter. The $\vect{q}$-resolved local density of states (LDOS) $\rho^{\mu}_{j}(E, \vect{q})$ on sublattice site $\mu$ on layer $j$ is then
\begin{equation} \label{eqn:ldos}
\rho^{\mu}_{j}(E,\vect{q})=-\frac{2}{\pi} \mathrm{Im} \left[ G^{\mu,\mu}_{j,j} (E + i \eta, \vect{q}) \right],
\end{equation}
where $E$ is the energy, $G^{\mu,\mu}_{j,j}=\left<\Psi^{\mu}_{j}\right| G \left|\Psi^{\mu}_{j}\right>$, the factor of two accounts for electron spin degeneracy, and $\eta$ ensures numerical stability at the cost of a Lorentzian broadening of spectral features of half width at half maximum of $\eta$. Unless otherwise stated, we use $\eta=1$ meV when calculating the LDOS. The site-resolved LDOS $\rho^{\mu}_{j}(E)$ is
\begin{equation} \label{eqn:dos}
\rho^{\mu}_{j} (E) = \frac{\Omega}{(2 \pi)^{2}} \int_{\textrm{BZ}} \rho^{\mu}_{j} (E, \vect{q}) \ d^2\vect{q},
\end{equation}
where $\Omega = \tfrac{\sqrt{3}}{2} a^{2}$ is the area of the in-plane unit cell, and BZ denotes the first Brillouin zone --- see Fig.\@ \ref{fig:structure}c. In this paper, we focus on low-energy electronic behavior which occurs in the vicinity of the valleys centered on $K$ and $K^{\prime}$. Results are presented for valley $K$, with those at $K^{\prime}$ related by time-inversion. The LDOS of bulk graphite of both stackings is given in Fig.\@ \ref{fig:bulkdos} to allow for comparison to the junction results in Section \ref{sec:ES}.

We project out the electronic properties on layers $j=1,\dots,N$, centered on the junction and termed the interface region. Layers $-\infty<j<1$ and $N<j<\infty$ are perfect rhombohedral or Bernal graphite left and right half-crystals.
The $2N\times 2N$ interfacial elements of the Green's function, Eq.\@ \ref{eqn:Gdefinition}, are obtained using
\begin{equation}\label{eqn:Gembed}
G(\omega,\vect{q})=\left[ \omega-{\mathcal H}(\vect{q})-\Sigma^{\textrm{L}}(\omega,\vect{q})-\Sigma^{\textrm{R}}(\omega,\vect{q})\right]^{-1},
\end{equation}
where ${\mathcal H}(\vect{q})$ is the $2N\times 2N$ Hamiltonian constructed in the same way as $H$ in Eq.\@ (\ref{eqn:hamiltonian}) but for a small number of layers in the interfacial region only, and $\Sigma^{\textrm{L}}$ ($\Sigma^{\textrm{R}}$) are embedding potentials accounting exactly for the influence of the left (right) half-crystal. In Eq.\@ (\ref{eqn:Gembed}), only the $j,j'=1,1$ block of $\Sigma^{\textrm{L}}$ and the $j,j'=N,N$ block of $\Sigma^{\textrm{R}}$ are non-zero; they are explicitly,
\begin{subequations}\label{eqn:Sigma}
\begin{equation}
\hat{\Sigma}^{\textrm{L}}_{1,1}(\omega,\vect{q})=\hat{V}^\dagger_{0,1}\hat{G}^{\textrm{L}}_{0,0}(\omega,\vect{q})\hat{V}_{0,1},
\end{equation}
\begin{equation}
\hat{\Sigma}^{\textrm{R}}_{N,N}(\omega,\vect{q})=\hat{V}_{N,N+1}\hat{G}^{\textrm{R}}_{N+1,N+1}(\omega,\vect{q})\hat{V}^\dagger_{N,N+1}.
\end{equation}
\end{subequations}
Here $\hat{G}_{0,0}^{\textrm{L}}$ and $\hat{G}_{N+1,N+1}^{\textrm{R}}$ are the $2\times 2$ surface blocks of the Green's functions for the half-crystals on the left and right. In numerical calculation we find these using decimation \cite{LannooFriedel}.

\subsection{\label{subsec:minimal}Minimal model}
For interpretation and analysis we make use of the simpler "minimal" model in which $\gamma_{3}=\gamma_{4}=\Delta^{\mu}_{j}=0$. In this regime, analytic expressions for $\hat{G}_{0,0}^{\textrm{L}}$ and $\hat{G}_{N+1,N+1}^{\textrm{R}}$ can be derived from Eq.\@ (\ref{eqn:Gembed}) by setting one of $\Sigma^{\textrm{L}}$ and $\Sigma^{\textrm{R}}$ to zero, and exploiting the fact that adding to a semi-infinite crystal a single layer (rhombohedral) or bilayer (Bernal) while maintaining the stacking sequence results in an equivalent crystal (see the Supplemental Material for details \cite{supplemental}. These are then used in Eq\@. (\ref{eqn:Sigma}) to obtain self-energies for the half-crystals to the left and right of layer $J$, which used in Eq\@. (\ref{eqn:Gembed}) with ${\mathcal H} = \hat{H}^{0}_{J}$ yield an analytic expression for the $2\times 2$ Green's function for layer $J$. The poles of this give the dispersion relation of junction states. If the Green's functions for other layers are required, $\Sigma^{\textrm{L}}$ and $\Sigma^{\textrm{R}}$ are modified to account for the particular stacking sequence either side of the layer in question.

\section{\label{sec:charge}Charge Redistribution}

The breaking of translational symmetry in the out-of-plane $z$ direction causes charge redistribution in the vicinity of the junction with an associated electrostatic potential. To model this, we consider $N$ parallel sheets of constant (non-zero) charge density with equal interlayer spacing $c$. The resulting electric potential $V(z)$ is
\begin{equation} \label{eqn:potential}
V(z) = - \frac{1}{2 \Omega \varepsilon_{d}} \sum_{j} \left( \bar{q}_{j} | z - z_{j} | + \bar{p}_{j} \text{sgn} \left(z - z_{j}\right) \right) + \alpha + \beta z,
\end{equation}
where $\bar{q}_{j}$ and $\bar{p}_{j}$ are the net charge and $\hat{z}$-dipole moment of a unit cell in layer $j$, and $\alpha$ and $\beta$ are constants fixed by the requirement that $V(z)$ is zero at the position of layer $j=0$ and $j=N+1$. We use $\varepsilon_{d} = 2.5 \varepsilon_{0}$ for the dielectric constant \cite{Tepliakov2021, Slizovskiy2021}. 

The excess charge on each layer is $\bar{q}_{j} = (2 - \xi_{j}) e$, with $e$ the elementary charge, and the number of electronic states $\xi_{j}$ obtained by integrating the LDOS over the occupied states,
\begin{equation} \label{eqn:electronstates}
\xi_{j} = \sum_{\mu=A,B} \int^{E_F}_{-\infty} \rho^{\mu}_{j}(E) \ d E.
\end{equation}
Here, we use a broadening of $\eta=10^{-6}$ meV when calculating the LDOS. The total dipole moment $\bar{p}_{j}$ associated with a given layer is calculated from \cite{Wu2017} 
\begin{equation} \label{eqn:dipole}
\bar{p}_{j} =e\sum_{\mu}\sum_{j',\mu'}M_{j,j'}^{\mu,\mu'}Z^{\mu,\mu'}_{j,j'},
\end{equation}
\begin{equation}\label{eqn:GF_ME}
M^{\mu,\mu'}_{j,j'}=-\frac{\Omega}{2\pi^{3}}\int^{E_F}_{-\infty}\!\!
\int_{BZ} \mathrm{Im} \left[ G^{\mu,\mu'}_{j,j'}(E + i \eta, \vect{q}) \right] \ d^2\vect{q}\ dE,
\end{equation}
\begin{equation} \label{eqn:dipoleintegral}
Z^{\mu,\mu'}_{j,j'} = \int z \psi(\vect{r}) \psi(\vect{r}+\vect{R}_{j}^{\mu}-\vect{R}_{j'}^{\mu'})d^{3}\vect{r},
\end{equation}
where $\psi$ is the carbon $2 p_{z}$ orbital, and $\vect{R}_{j}^{\mu}$ the location of sublattice site $\mu$ in layer $j$. Using Slater-type orbitals \cite{Slater1930} with effective nuclear charge 3.25 and atomic radius 0.65 \AA, the dominant contributions to $p_{j}$ come from $j'=j\pm 1$ with $Z^{\mu,\mu'}_{j,j'}=\pm 0.148$ \AA\  for $\vect{R}_{j}^{\mu}-\vect{R}_{j'}^{\mu'}=(0,0,\pm c)$ and $\pm 0.082$ \AA\ for $\vect{R}_{j}^{\mu}-\vect{R}_{j'}^{\mu'}=(\delta_x,\delta_y,\pm c)$ with $|\vect{\delta}|=a/\sqrt{3}$. Using these we find the contribution from the dipoles in Eq.\@ (\ref{eqn:potential}) is several orders of magnitude smaller than that from the charges.

\begin{figure}[t]
\centering
\includegraphics[width=0.95\columnwidth]{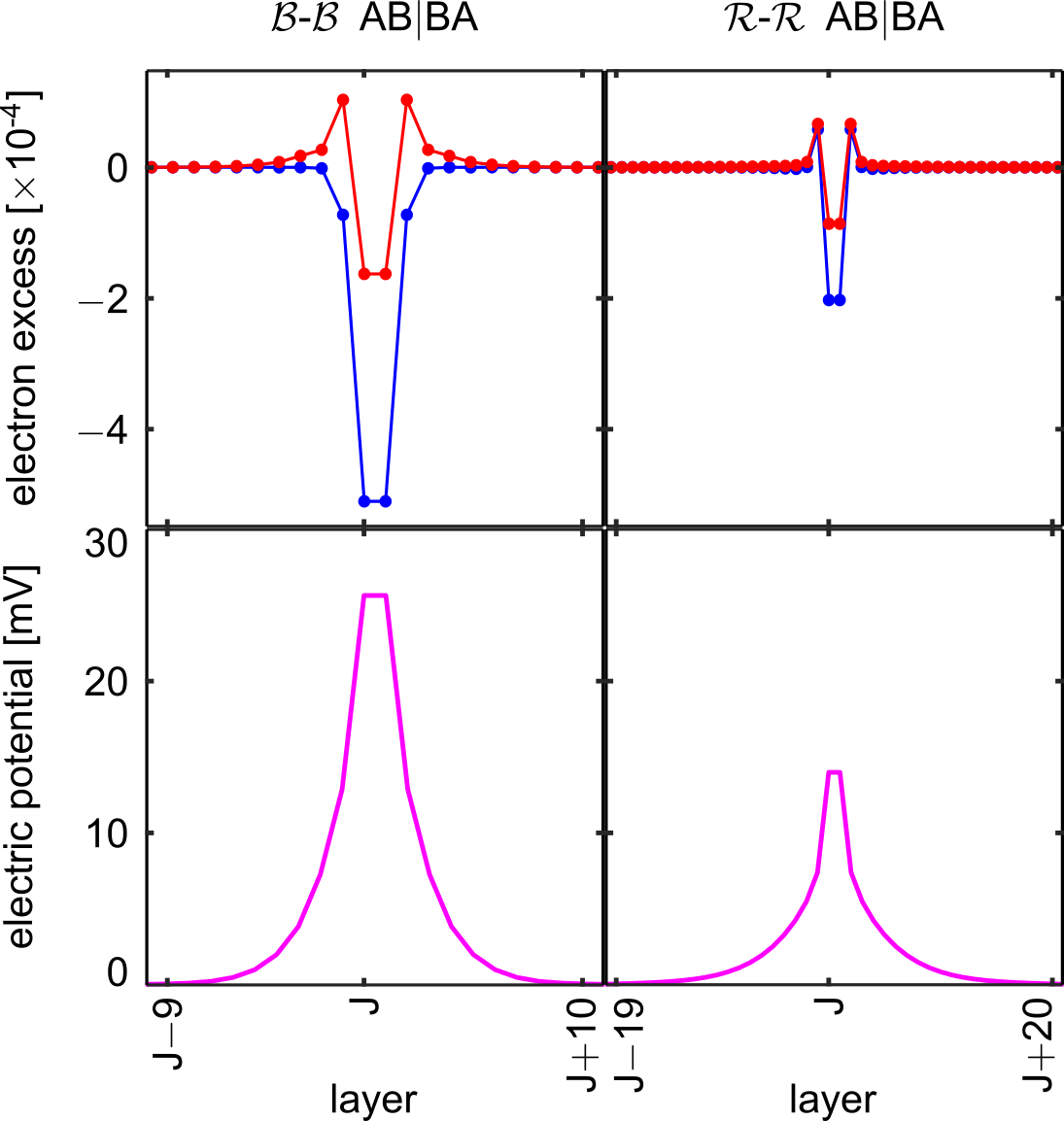}
\caption{\label{fig:chargeneutrality} The calculated electron excess and electric potential found in the vicinity of two example junctions: $\mathcal{B}$-$\mathcal{B}$ AB$|$BA and $\mathcal{R}$-$\mathcal{R}$ AB$|$BA. Red (blue) are the number of excess electrons on each layer after (before) the self-consistent procedure, and electric potential is shown in magenta.}
\end{figure}

The calculated electric potential is included in the Hamiltonian as an additional on-site energy shift $\Delta^{\mu}_{j} \rightarrow \Delta^{\mu}_{j} - V(|\vect{R}_{j}^{\mu}|)$. These change the resulting charges and dipoles, and so are calculated self-consistently using an interface region sufficiently large to ensure overall charge neutrality better than $10^{-6}e$. This typically requires 10 layers of Bernal or 20 layers of rhombohedral graphite on either side of the physical interface. The resulting on-site energy shifts are on the 10 meV scale, being largest at  $\mathcal{B}$-$\mathcal{B}$ AB$|$AC (reaching $30$ meV). 
Results for two example junctions are given in 
Fig.\@ \ref{fig:chargeneutrality}.

\section{\label{sec:ES}Electronic structure}

\begin{figure*}[t]
\centering
\includegraphics[width=\textwidth]{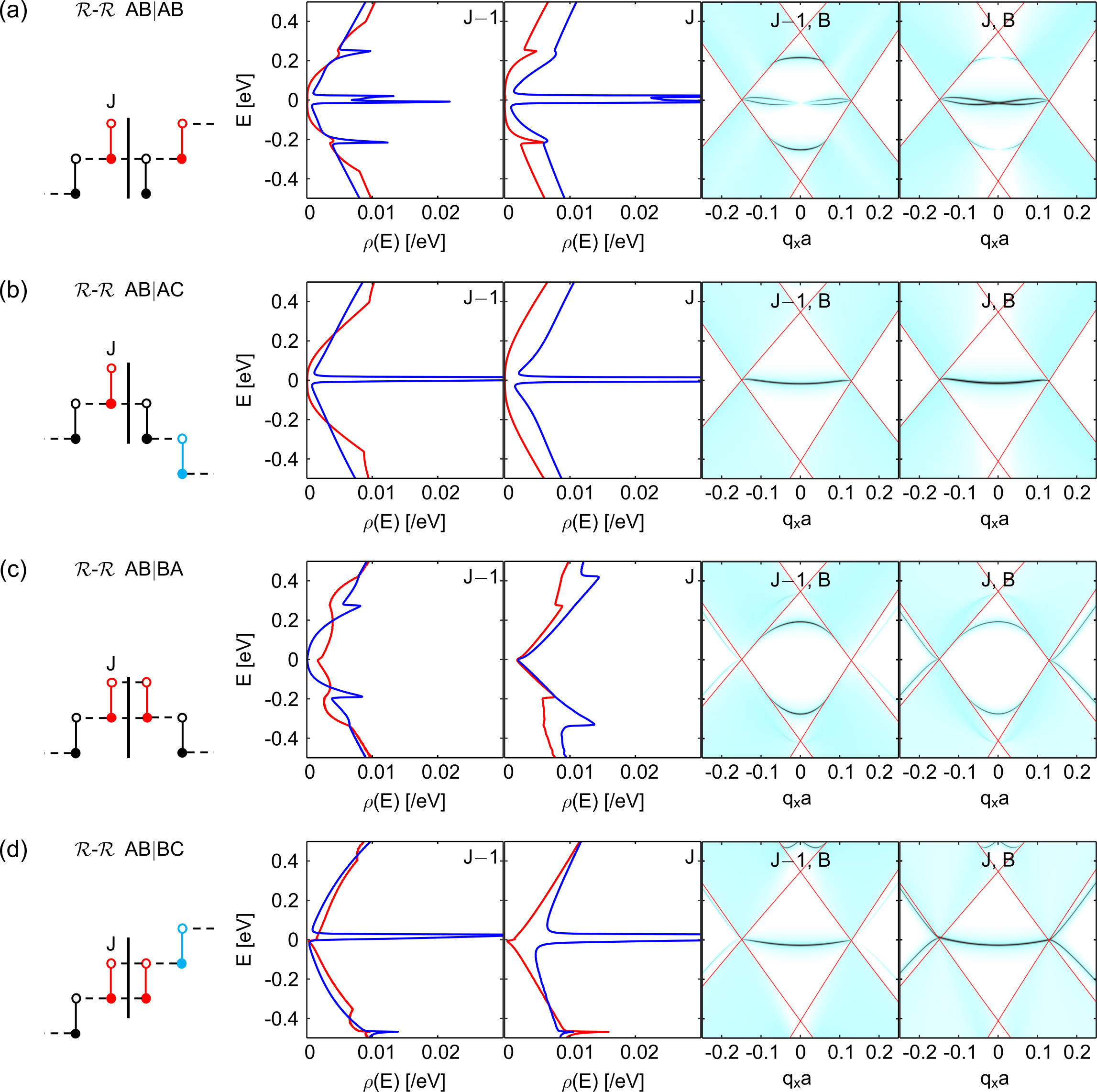}
\caption{\label{fig:rrdosresults} The site-resolved and $\vect{q}$-resolved density of states at the junction of the a) $\mathcal{R}$-$\mathcal{R}$ AB$|$AB, b) $\mathcal{R}$-$\mathcal{R}$ AB$|$AC, c) $\mathcal{R}$-$\mathcal{R}$ AB$|$BA, and d) $\mathcal{R}$-$\mathcal{R}$ AB$|$BC crystals. These results use the same color scheme as described in the caption of Fig.\@ \ref{fig:bulkdos}.}
\end{figure*}

\begin{figure*}[t]
\centering
\includegraphics[width=\textwidth]{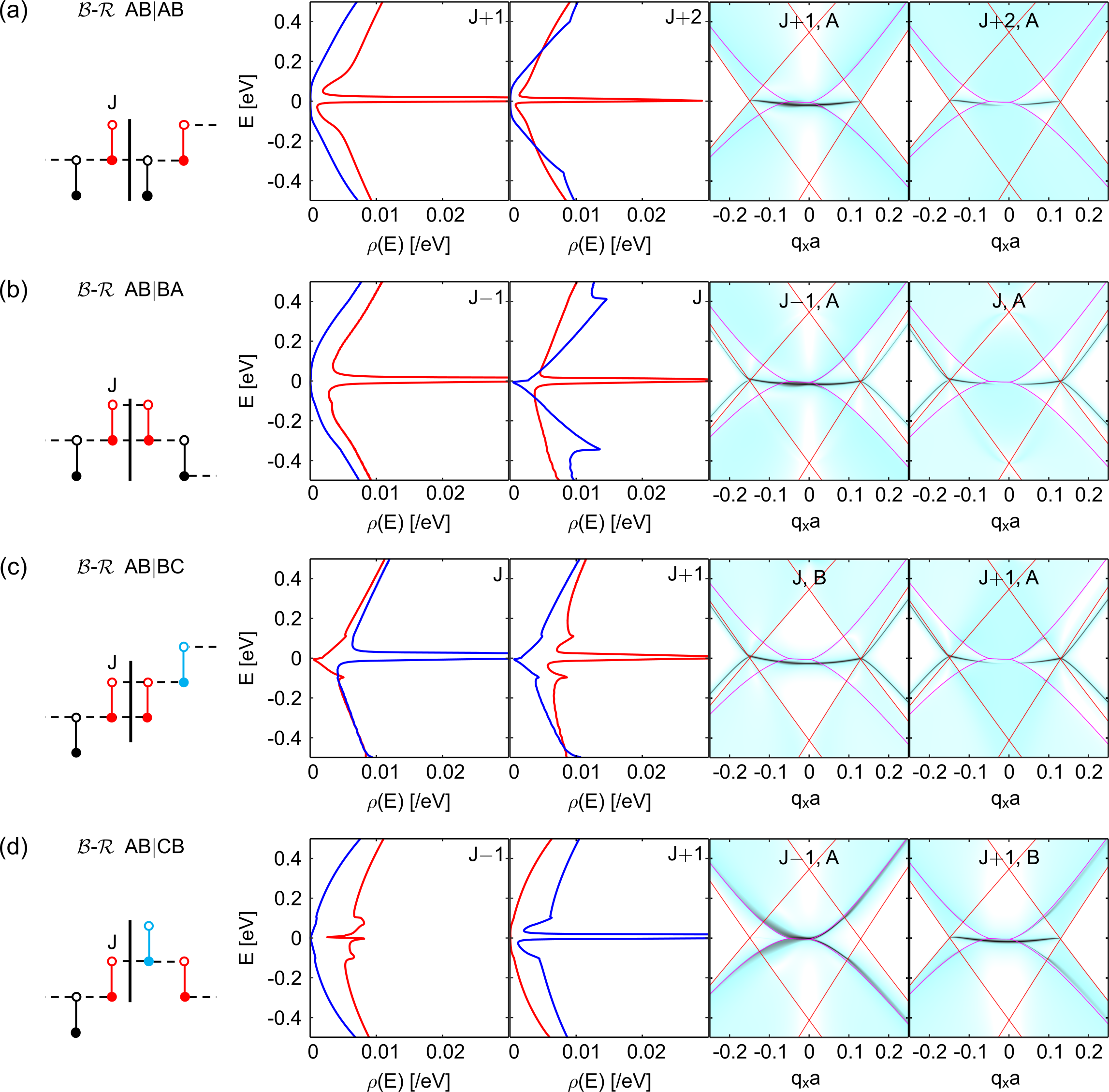}
\caption{\label{fig:brdosresults} The site-resolved and $\vect{q}$-resolved density of states at the junction of the a) $\mathcal{B}$-$\mathcal{R}$ AB$|$AB, b) $\mathcal{B}$-$\mathcal{R}$ AB$|$BA, c) $\mathcal{B}$-$\mathcal{R}$ AB$|$BC, and d) $\mathcal{B}$-$\mathcal{R}$ AB$|$CB crystals. These results use the same color scheme as described in the caption of Fig.\@ \ref{fig:bulkdos}.}
\end{figure*}

\begin{figure*}[t]
\centering
\includegraphics[width=\textwidth]{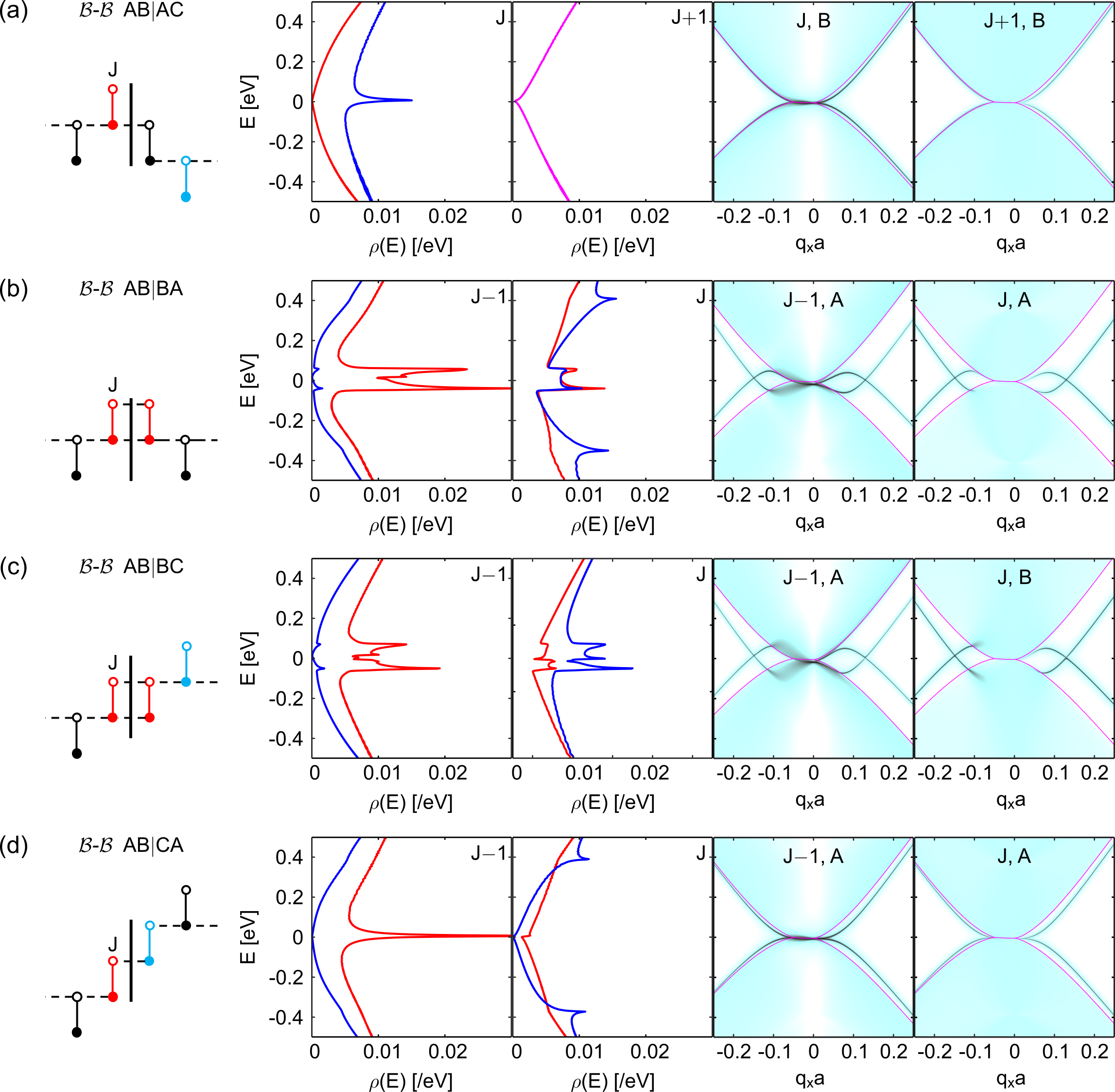}
\caption{\label{fig:bbdosresults} The site-resolved and $\vect{q}$-resolved density of states at the junction of the a) $\mathcal{B}$-$\mathcal{B}$ AB$|$AC, b) $\mathcal{B}$-$\mathcal{B}$ AB$|$BA, c) $\mathcal{B}$-$\mathcal{B}$ AB$|$BC, and d) $\mathcal{B}$-$\mathcal{B}$ AB$|$CA crystals. These results use the same color scheme as described in the caption of Fig.\@ \ref{fig:bulkdos}.}
\end{figure*}

We probe the electronic behavior at the junctions by studying the low-energy LDOS. Key results for each junction are shown in Figs.\@ \ref{fig:rrdosresults}, \ref{fig:brdosresults} and \ref{fig:bbdosresults}, with full results for all eight sites in the four-layer junction regions reported in the Supplemental Material \cite{supplemental}. The $\vect{q}$-resolved LDOS on sites at the junctions show states occupying the region of energy-wave vector space corresponding to the continua of the two semi-infinite crystals to either side of the junction, along with discrete bands of junction-localized states that lie outside of these continua. We find junction-localized states are present in all 12 crystals.

\begin{figure*}[t]
\centering
\includegraphics[width=0.95\textwidth]{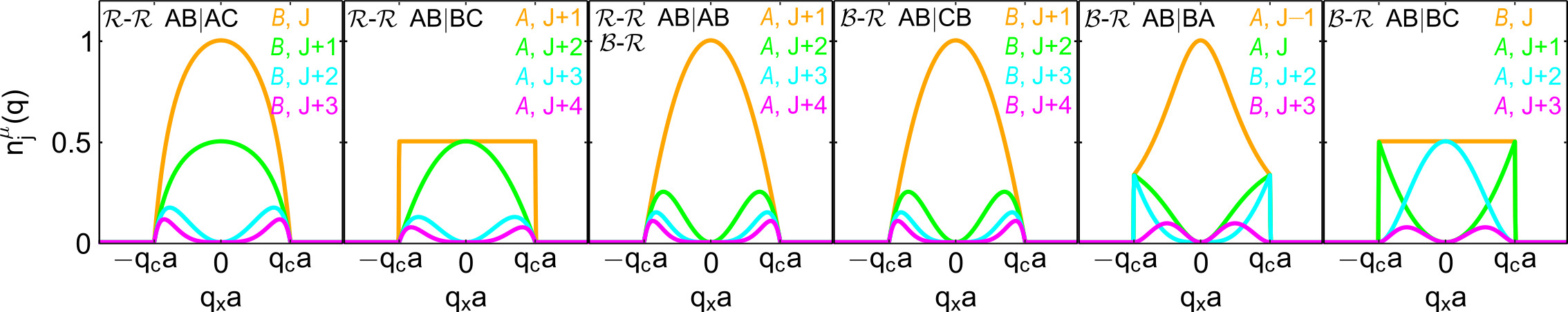}
\caption{\label{fig:distribution} The number of zero-energy states (per spin per valley) $n_{j}^{\mu}$ on sublattice $\mu$ in near-junction layer $j$ at all junctions possessing such states, as a function of wave vector $q_x$ for $q_y=0$. Calculated in the minimal model. The critical wave vector $q_c$ satisfying $|f_{\vect{q}_c}|=\gamma_{1}/\gamma_{0}$ marks the boundary between the topologically trivial and non-trivial regimes.}
\end{figure*}

Results for the $\mathcal{R}$-$\mathcal{R}$ crystals are shown in Fig.\@ \ref{fig:rrdosresults}. Most strikingly, in three cases there are weakly-dispersive junction-localized states that exist on one sublattice only, lying near the Fermi energy ($E_F=0$) for a finite range of wave vectors and which originate a prominent peak in the site-resolved LDOS. This hints at a topological origin \cite{Soneji2026}. Recall the SSH model \cite{Su1979, Asboth2016}, a one-dimensional chain of sites connected by alternating hopping parameters (intra-cell and inter-cell). In this model topologically non-trivial electronic states emerge on the edge sites of an extended chain when the inter-cell coupling is stronger than the intra-cell coupling. In the absence of higher-order interactions they are pinned at $E=0$ and robust to perturbations. It has previously been recognized \cite{Guinea2006, Manes2007, Muten2021, Soneji2026} that rhombohedral graphite provides a realization of the SSH model with intra-cell coupling $\gamma_{0}f_{\vect{q}}$ and inter-cell coupling $\gamma_{1}$, so that when $\gamma_{0}|f_{\vect{q}}|<\gamma_{1}$, one is in a topologically non-trivial regime resulting in the presence of zero-energy edge states at the surface of a rhombohedral stack, with amplitude on one sublattice only. This condition is satisfied at wave vectors $|\vect{q}| < |\vect{q}_{c}|$, i.\@e.\@ between the touching points of the continua --- see Fig.\@ \ref{fig:bulkdos}b. Broken symmetry following the inclusion of additional couplings ($\gamma_{4}$) causes small energy shifts of these edge states \cite{Koshino2009} which continue to present as flat bands. Junction states in Fig.\@ \ref{fig:rrdosresults} exhibit similar behavior to such edge states.  Mapping on to the SSH model means that, before alignment, the two constituent rhombohedral half-crystals possess one zero-energy state each (per electron spin) at wave vectors in the vicinity of $K$. Accounting for the electronic spin degeneracy, as well as the corresponding electronic states at valley $K^{\prime}$, means there are a total of eight junction states at the $\mathcal{R}$-$\mathcal{R}$ junctions for each $\vect{q}$ in the topologically non-trivial regime \cite{Soneji2026}. Using the minimal model (Sec.\@ \ref{subsec:minimal}), we find their spatial distribution in terms of the number of zero-energy topological states $n_{j}^{\mu}(\vect{q})$ (per spin per valley) on a particular site belonging to sublattice $\mu$ in layer $j$. For example, for the $\mathcal{R}$-$\mathcal{R}$ AB$|$AC junction, when $\gamma_{0}|f_{\vect{q}}|<\gamma_{1}$,

\begin{equation} \label{eqn:nRRABAC}
\begin{split}
n^{B}_{J}(\vect{q}) &= \frac{2\left(\gamma_{1}^{2}-\gamma_{0}^{2}|f_{\vect{q}}|^{2}\right)}{2\gamma_{1}^{2}-\gamma_{0}^{2}|f_{\vect{q}}|^{2}}, \\
n^{B}_{J \pm m}(\vect{q}) &= \frac{\left(\gamma_{0}^{2}|f_{\vect{q}}|^{2}\right)^{m-1} \left(\gamma_{1}^{2}-\gamma_{0}^{2}|f_{\vect{q}}|^{2}\right)}{\gamma_{1}^{2(m-1)} \left(2\gamma_{1}^{2}-\gamma_{0}^{2}|f_{\vect{q}}|^{2}\right)}, \ \ m \geq 1,
\end{split}
\end{equation}
and $n_{j}^{A}=0$ for all $j$. The distributions in Eqn.\@ (\ref{eqn:nRRABAC}) are shown in the first panel in Fig.\@ \ref{fig:distribution}. In this crystal zero-energy states occupy $B$ sublattice sites only. At $\vect{q}=0$ (when $f_{\vect{q}}=0$) they are localized on layers $J$ and $J \pm 1$ only, on formation of the junction, with layer $J$ hosting four states and layers $J\pm 1$ two states each. As $|\vect{q}|$ increases, the states become increasingly delocalized, extending further away from the junction. This particular crystal, corresponding to a twin boundary, has been studied previously \cite{Muten2021, GarciaRuiz2023}. Expressions resulting from a similar analysis of the spatial distribution of topological zero-energy states at the other $\mathcal{R}$-$\mathcal{R}$ junctions can be found in the Supplemental Material \cite{supplemental}, and are plotted in the firs three panels of Fig.\@ \ref{fig:distribution}. For the $\mathcal{R}$-$\mathcal{R}$ AB$|$BA junction there is no zero-energy flat band, Fig.\@ \ref{fig:rrdosresults}c. For this geometry there is no symmetry-protection of the edge states of the two half-crystals when they are brought together, which instead hybridize strongly \cite{Soneji2026} and move away from $E=0$ into the continuum; note the in-gap states seen for $|\vect{q}|<|\vect{q}_c|$ are not the result of this hybridization, but originate from high-energy states that move \textit{towards} $E=0$ as the half-crystals come together. Unlike the topological states, these states are seen to extend over both sublattices (Supplemental Material Fig.\@ 6 \cite{supplemental}). Analytic expressions for their dispersion, and the dispersion relations of other dispersive junction states at the $\mathcal{R}$-$\mathcal{R}$ junctions, are given in the Supplemental Material \cite{supplemental}.

In the case of $\mathcal{B}$-$\mathcal{R}$ crystals, Fig.\@ \ref{fig:brdosresults}, only the right half-crystal is rhombohedral stacked, possessing one topological zero-energy edge state (per spin, per valley) for wave vectors $|\vect{q}<|\vect{q}_{c}|$. The left Bernal half-crystal has no such state, and so hybridization of rhombohedral edge states as happens at $\mathcal{R}$-$\mathcal{R}$ AB$|$BA cannot occur. We find weak coupling to the Bernal continuum is present, but a clearly identifiable weakly dispersive band near $E=0$ exists at all four of these junctions, extending over wave vectors corresponding to the topologically non-trivial regime of the rhombohedral half-crystal. Using the minimal model, at $\vect{q}=0$ and for the $\mathcal{B}$-$\mathcal{R}$ AB$|$AB and $\mathcal{B}$-$\mathcal{R}$ AB$|$CB junctions we find that the states are entirely localized on the edge atom of the semi-infinite rhombohedral region, extending further into the rhombohedral stacked side of the junction with increasing $\vect{q}$. For the former, this localization is the same as that of one of the zero-energy states in the $\mathcal{R}$-$\mathcal{R}$ AB$|$AB junction; for the latter the behavior is the same but on the other sublattice. In turn, for the $\mathcal{B}$-$\mathcal{R}$ AB$|$BA junction the spatial distributions are,
\begin{equation} \label{eqn:nBRABBA}
\begin{split}
n^{A}_{J-1}(\vect{q}) &= \frac{\gamma_{1}^{2}}{\gamma_{1}^{2}+2\gamma_{0}^{2}|f_{\vect{q}}|^{2}}, \\
n^{A}_{J}(\vect{q}) &= \frac{\gamma_{0}^{2}|f_{\vect{q}}|^{2}}{\gamma_{1}^{2}+2\gamma_{0}^{2}|f_{\vect{q}}|^{2}}, \\
n^{B}_{J+1}(\vect{q}) &= \frac{\gamma_{0}^{4}|f_{\vect{q}}|^{4}}{\gamma_{1}^{2}\left(\gamma_{1}^{2}+2\gamma_{0}^{2}|f_{\vect{q}}|^{2}\right)}, \\
n^{B}_{J+m}(\vect{q}) &= \frac{\left(\gamma_{0}^{2}|f_{\vect{q}}|^{2}\right)^{2(m-1)}\left(\gamma_{1}^{2}-\gamma_{0}^{2}|f_{\vect{q}}|^{2}\right)^{2}}{\gamma_{1}^{2m}\left(\gamma_{1}^{2}+2\gamma_{0}^{2}|f_{\vect{q}}|^{2}\right)}, \ \ m \geq 2,
\end{split}
\end{equation}
shown in Fig.\@ \ref{fig:distribution}, and $n_{j}^{\mu}=0$ on all other atoms. Notably, 
at $\vect{q}=0$ (when $f_{\vect{q}}=0$), the zero-energy state is entirely localized on layer $J-1$, which lies within the Bernal stacked layers and two layers away from the edge of the rhombohedral half-crystal. At this wave vector the only non-zero elements of the Hamiltonian are the $\gamma_{1}$ elements, which cause the sites in the rhombohedral region to undergo interlayer dimerization, half of the sites in the Bernal region to decouple from those on the backbone, and the sites of the BB bilayer at the junction to either dimerize across the junction or join the backbone. Zero-energy states cannot be hosted by the dimers or the semi-infinite linear chain formed by the backbone, so the junction state emerges on the closest uncoupled atom from the junction: the $A$ sublattice atom in layer $J-1$. As $\vect{q}$ increases, in-plane hopping is recovered and the junction states extend into the rhombohedral region. Similar behavior is also observed at the $\mathcal{B}$-$\mathcal{R}$ AB$|$BC junction --- see the Supplemental Material for expressions for the number of zero-energy states at the other $\mathcal{B}$-$\mathcal{R}$ junctions which are plotted in Fig.\@ \ref{fig:distribution} \cite{supplemental}. Dispersive states localized at the junction are also observed in the $\mathcal{B}$-$\mathcal{R}$ AB$|$BA, $\mathcal{B}$-$\mathcal{R}$ AB$|$BC, and $\mathcal{B}$-$\mathcal{R}$ AB$|$CB crystals, but we find that no simple analytical expressions for their dispersion relations exist. To note, the $\mathcal{B}$-$\mathcal{R}$ AB$|$CB junction can also be viewed as containing a finite reversed rhombohedral trilayer, consisting of layers $J-1$, $J$ and $J+1$, sandwiched between the semi-infinite Bernal and rhombohedral half-crystals. However, layer $J+1$ is also the edge layer of the right rhombohedral half-crystal. Low-energy effective theory \cite{Min2008} shows that in finite rhombohedral stacks of $M+L-1$ layers with stacking reversing at layer $M$ the system decomposes into two subsystems with opposite chirality of lengths $\max(M,L)$ and $\min(M-1,L-1)$. Here $M=3$ and $L\rightarrow\infty$, and so the finite reversed section behaves like a $N=2$ $N$-layer rhombohedral film. In the minimal model this supports states dispersing like $|E|\sim q^2$ and site-resolved LDOS $\rho^{j}_{\mu}(E)\sim |E|^{0}$, which explains the absence of a prominent peak on sites in layer $J-1$, the left edge of the reversed rhombohedral segment (Fig.\@ \ref{fig:brdosresults}d). Further results on systems with increasing thickness of the reversed rhombohedral section between the Bernal and rhombohedral half-crystals confirm this, showing the development of an increasingly flat junction band and associated prominent LDOS peak (see Supplemental Material \cite{supplemental}) as the number of reversed rhombohedral layers grows.

Results for the $\mathcal{B}$-$\mathcal{B}$ crystals are shown in Fig.\@ \ref{fig:bbdosresults}. At the AB$|$BA and AB$|$BC junctions with AA-type stacking regions, Figs.\@ \ref{fig:bbdosresults}b and \ref{fig:bbdosresults}c, junction states outside of the bulk continuum that exhibit non-monotonic multi-extremal dispersion, effectively forming a superposition of W-shaped and inverted W-shaped profiles, cause a significant enhancement of the density of states in the junction region over a $100-200$ meV range centered on the Fermi energy. In contrast, the LDOS at $\mathcal{B}$-$\mathcal{B}$ AB$|$AC and AB$|$CA junctions exhibits a small narrow peak near $E=0$ on particular sites, Figs.\@ \ref{fig:bbdosresults}a and \ref{fig:bbdosresults}d. Alignment of the half-crystals when forming these junctions results in three- and four-layer rhombohedral stacked regions respectively. Within the minimal model we find that the AB$|$AC junction possesses discrete states that disperse as \begin{equation} \label{eqn:dispersionABAC}
|E| = \frac{\gamma_{1}^{2}+2\gamma_{0}^{2}|f_{\vect{q}}|^{2}-\gamma_{1}\sqrt{\gamma_{1}^{2}+4\gamma_{0}^{2}|f_{\vect{q}}|^{2}}}{2 \gamma_{0}|f_{\vect{q}}|},
\end{equation}
whilst at the  AB$|$CA junction states exist satisfying
\begin{equation} \label{eqn:dispersionABCA}
2\gamma_{0}^2|f_{\vect{q}}|^{2}=\left( |E|+\sqrt{4\gamma_{1}|E|+|E|^2}\right)\left( \gamma_{1}+|E|\right).
\end{equation}
For small $q$ these dispersions are $|E|\simeq \gamma_{0}^{3}|f_{\vect{q}}|^{3}/\gamma_{1}^{2}\sim q^{3}$ and $|E|\simeq \gamma_{0}^{4}|f_{\vect{q}}|^{4}/\gamma_{1}^{3}\sim q^{4}$, consistent with the $|E|\sim q^{N}$ dispersion of states in $N$-layer rhombohedral graphite thin films \cite{Min2008} that are the finite-thickness precursors of the bulk topological edge states formed in thick rhombohedral graphite. The corresponding site-resolved LDOS $\rho^{j}_{\mu}(E)\sim |E|^{(2-N)/N}$ in the finite $N$-layer films, which diverges as $|E|\rightarrow 0$, and is restricted to sites on opposing sublattices at each end of the rhombohedral film. This is consistent with where we find spectral weight lies at the $\mathcal{B}$-$\mathcal{B}$ junctions: $(j,\mu)=(J, B)$ and $(J+2,A)$ at AB$|$AC, $(J-1,A)$ and $(J+2,B)$ at AB$|$CA.

\section{\label{sec:discussion}Discussion}

Our results provide a comprehensive study of the electronic behavior in 12 distinct aperiodic crystals formed by the embedding of one of six possible junctions between the surfaces of two semi-infinite graphite crystals of Bernal, rhombohedral, or mixed stacking. We find localized states to be a ubiquitous feature. Most striking are the localized states forming flat bands at wave vectors close to the $K$ and $K'$ points of the Brillouin zone, in rhombohedral-rhombohedral and Bernal-rhombohedral systems. These derive from the topological edge states at the surface of a rhombohedral half-crystal, and which, apart from the $\mathcal{R}$-$\mathcal{R}$ AB$|$BA case, survive the coupling to the second half-crystal. 

\begin{table}[t]
\caption{Band widths and localization of topological bands at graphite junctions in Figs.\@ \ref{fig:rrdosresults} and \ref{fig:brdosresults}. $\Delta\omega$ ($\Delta\omega_{0}$): calculated band widths with (without) charge self-consistency. $\Gamma$: resonance half-width within Bernal continuum (estimated at $q_{x}a=-0.05$); $f$: fraction of the associated state density within the region consisting of layers $J-j_{\text{min}}$ to $J+j_{\text{max}}$, calculated within the minimal model. The final column details the range $j_{\text{min}}..j_{\text{max}}$ }
\label{tab:widths}
\begin{ruledtabular}
\begin{tabular}{cccccc}
Junction & $\Delta\omega$ & $\Delta\omega_{0}$ & $\Gamma$ & $f$ & region\\
& (meV) & (meV) & (meV) &  & \\
\hline
${\mathcal R}$-${\mathcal R}$\ AB$|$AB  & 36 & 42 & & 0.75 & -2..3\\
${\mathcal R}$-${\mathcal R}$\ AB$|$AC  & 22 & 34 & & 0.79 & -3..3\\
${\mathcal R}$-${\mathcal R}$\ AB$|$BC  & 43 & 43 & & 0.83 & -2..3\\
${\mathcal B}$-${\mathcal R}$\ AB$|$AB  & 24 & 35 & 3.4 & 0.86 & 1..6\\
${\mathcal B}$-${\mathcal R}$\ AB$|$BA  & 28 & 39 & 3.0 & 0.98 & -1..3\\
${\mathcal B}$-${\mathcal R}$\ AB$|$BC  & 41 & 44 & 1.6 & 0.98 & 0..5\\
${\mathcal B}$-${\mathcal R}$\ AB$|$CB  & 23 & 34 & 0.6 & 0.86 & 1..6\\
\end{tabular}
\end{ruledtabular}
\end{table}

In Table \ref{tab:widths} we list some properties of these bands. The band widths $\Delta \omega$ are small, ranging from $23$ to $43$ meV, comparable to the $\sim 30$ meV width of the corresponding band observed at the surface of rhombohedral graphite crystals using scanning-tunneling \cite{Pierucci2015, Hagymasi2022} and angle-resolved photoemission \cite{Zhang2024} spectroscopies. Interestingly, the inclusion of charge self-consistency, which results from the broken translational symmetry in the direction perpendicular to the plane of the junction, leads to on-site energy shifts up to $\sim 10-20$ meV decaying over $5-10$ layers from the junction, and which act to narrow the width, irrespective of whether the charge transfer makes the junction more or less attractive to electrons. This reflects a complex interplay between the spatial distributions of the electric potential and of the flat-band states, which causes a wave vector-dependent shift in their energies. The narrowing due to charge self-consistency is significant: over 25\% in four of the systems. When a Bernal half-crystal is involved, the flat band partially overlaps with the Bernal continuum, but still retains its integrity. The resulting resonance has a half-width a fraction of the overall band width.

The states within the band are strongly confined to the vicinity of the junction. To quantify this, we integrate the wave vector resolved distributions $n_{j}^{\mu}(\vect{q})$ obtained in the minimal model, Fig.\@ \ref{fig:distribution} and Supplemental Material \cite{supplemental}, over wave vectors $|\vect{q}|<q_{c}$, and then sum them over sublattice sites to obtain layer-resolved charge counts. From these, we find a fraction $f \ge 0.75$ of the charge associated with the band lies within a region spanning $n_r=5-7$ layers, depending upon the system (Table \ref{tab:widths}). Including  spin-, valley-, and edge-state degeneracy $g$, the resulting density of flat-band states
\begin{equation}
n\simeq\frac{\pi q_{c}^2}{(2\pi)^2}\frac{f}{n_r c_0} g = \frac{4}{3\pi}\frac{\gamma_{1}^{2}}{\gamma_{0}^{2}}\frac{f}{a^2 c_0\, n_r} \times \left\{ \begin{array}{ll}1 \qquad & {\mathcal B}\text{-}{\mathcal R} \\ 2 & {\mathcal R}\text{-}{\mathcal R} \end{array} \right.
\end{equation}
indicates an average electron separation $d\simeq n^{-1/3}=2-3$ nm, and electron-electron interaction $U\simeq {e^2}/({4\pi\epsilon d})\sim 100-200$ meV, much greater than the band width $\Delta\omega$. The dramatically enhanced density of states at the Fermi energy and $U\gg\Delta\omega$, indicate electronic instabilities and strong correlation effects are to be expected, similarly to the case of flat bands at the surfaces of rhombohedral graphite \cite{Otani2010, Xu2012, Munoz2013, Kopnin2013, Henck2018, Hagymasi2022, Awoga2023, Zhang2024}.

In the case of $\mathcal{B}$-$\mathcal{B}$ junctions, neither of the Bernal half-crystals supports a topological edge state that can contribute to the junction electronic structure. Nevertheless, we find a marked enhancement of the low-energy density of states when the layer stacking at the junction presents a finite rhombohedral-stacked sequence, Fig.\@ \ref{fig:bbdosresults}a,d, albeit less pronounced than that due to the flat-bands at junctions involving rhombohedral half-crystal(s). This enhancement originates from junction-localized bands that disperse similar to the nascent flat-band states arising in rhombohedral tri-layer \cite{Coletti2013, Bao2017} and tetra-layer \cite{wang2018} rhombohedral graphene films. Rhombohedral trilayer graphene is also a strongly-correlated electron system, known to exhibit unconventional superconductivity \cite{Zhou2021, Pantaleon2023, Bostrom2024, Han2025}.

Thus, our results show graphite junctions warrant further investigation for exploring and understanding interaction effects in intrinsic carbon systems. At both $\mathcal{R}$-$\mathcal{R}$ and $\mathcal{B}$-$\mathcal{R}$ junctions, we find alignments with BB stacking that also exhibit low-energy flat-band electronic structure, suggesting twisted junctions where such configurations occur locally are also systems likely to exhibit novel electronic structures.

\section*{Acknowledgments}
This work has been supported by the UK Engineering and Physical Sciences Research Council (EPSRC) through grant EP/W524712/1.

\section*{Data availability}
All the data supporting the findings of this study are available within the paper and its Supplemental Material file or can be obtained using the equations therein. 

\bibliography{paper}

\end{document}


\title{Supplemental Material: \\ Electron localization, charge redistribution, and emergence of topological states at graphite junctions from Green's function embedding}

\author{L. Soneji}
\affiliation{Department of Physics, University of Bath, Claverton Down, Bath, BA2 7AY, United Kingdom}
\author{S. Crampin}
\affiliation{Department of Physics, University of Bath, Claverton Down, Bath, BA2 7AY, United Kingdom}
\author{M. Mucha-Kruczy\'{n}ski}
\email{M.Mucha-Kruczynski@bath.ac.uk}
\affiliation{Department of Physics, University of Bath, Claverton Down, Bath, BA2 7AY, United Kingdom}

\date{\today}
\maketitle

\tableofcontents

\newpage
\section{Analytic surface Green's function derivation}

Adding a single layer to a semi-infinite crystal of rhombohedral graphite while maintaining the stacking sequence results in a equivalent semi-infinite rhombohedral graphite crystal. Applying Eqns.\@ 8 and 9 from the main paper to an (ABC)-stacked rhombohedral half-crystal to the left of a single layer yields
\begin{equation}
\hat{G}^{\textrm{L}}_{0,0} (\omega, \vect{q}) = \left[ \omega - \hat{H}_{0} (\vect{q}) - \hat{V}_{\textrm{AB}}^{\dagger} \hat{G}^{\textrm{L}}_{0,0} (\omega, \vect{q}) \hat{V}_{\textrm{AB}} \right]^{-1},
\end{equation}
with solution
\begin{equation}
\hat{G}^{\textrm{L,(ABC)}}_{0,0} (\omega, \vect{q}) = \frac{\sigma_{\textrm{R}}}{\omega - \gamma_{1}^{2} \sigma_{\textrm{R}}}
\begin{pmatrix}
\omega & -\gamma_{0}f_{\vect{q}} \\
-\gamma_{0}f^{*}_{\vect{q}} & \omega - \gamma_{1}^{2} \sigma_{\textrm{R}}
\end{pmatrix},
\end{equation}
where
\begin{equation}
\sigma_{\textrm{R}} = \frac{\omega^{2} + \gamma_{1}^{2} - \gamma_{0}^{2} |f_{\vect{q}}|^{2} + \sqrt{\left( \omega^{2} + \gamma_{1}^{2} - \gamma_{0}^{2} |f_{\vect{q}}|^{2} \right)^{2} - 4 \gamma_{1}^{2} \omega^{2}}}{2 \gamma_{1}^{2} \omega}.
\end{equation}
For an (ABC)-stacked half-crystal to the right of a single layer
\begin{equation}
\hat{G}^{\textrm{R,(ABC)}}_{N+1,N+1} (\omega, \vect{q}) = \frac{\sigma_{\textrm{R}}}{\omega - \gamma_{1}^{2} \sigma_{\textrm{R}}}
\begin{pmatrix}
\omega - \gamma_{1}^{2} \sigma_{\textrm{R}} & -\gamma_{0}f_{\vect{q}} \\
-\gamma_{0}f^{*}_{\vect{q}} & \omega
\end{pmatrix}.
\end{equation}
Similar analysis for a (CBA)-stacked half-crystal yields $\hat{G}^{\textrm{L,(CBA)}}_{0,0}=\hat{G}^{\textrm{R,(ABC)}}_{N+1,N+1}$ and $\hat{G}^{\textrm{R,(CBA)}}_{N+1,N+1}=\hat{G}^{\textrm{L,(ABC)}}_{0,0}$.

Two layers must be added to an (AB)-stacked Bernal half-crystal in order to recover an equivalent crystal. This yields:
\begin{equation}
\begin{split}
\hat{G}^{\textrm{L,(AB)}}_{0,0} (\omega, \vect{q}) &=
\begin{pmatrix}
 \sigma_{\textrm{B}} & -\displaystyle\frac{\gamma_{0}f_{\vect{q}}\sigma_{\textrm{B}}}{\omega} \\
-\displaystyle\frac{\gamma_{0}f^{*}_{\vect{q}}\sigma_{\textrm{B}}}{\omega} & \displaystyle\frac{\omega+\gamma_{0}^{2}|f_{\vect{q}}|^{2}\sigma_{\textrm{B}}}{\omega^{2}}
\end{pmatrix}, \\
\hat{G}^{\textrm{R,(AB)}}_{N+1,N+1} (\omega, \vect{q}) &=
\begin{pmatrix}
\displaystyle\frac{\omega+\gamma_{0}^{2}|f_{\vect{q}}|^{2}\sigma_{\textrm{B}}}{\omega^{2}} & -\displaystyle\frac{\gamma_{0}f_{\vect{q}}\sigma_{\textrm{B}}}{\omega} \\
-\displaystyle\frac{\gamma_{0}f^{*}_{\vect{q}}\sigma_{\textrm{B}}}{\omega} & \sigma_{\textrm{B}}
\end{pmatrix},
\end{split}
\end{equation}
where
\begin{equation}
\sigma_{\textrm{B}} = \frac{\omega^{2} - \gamma_{0}^{2} |f_{\vect{q}}|^{2} + \sqrt{\left( \omega^{2} - \gamma_{0}^{2} |f_{\vect{q}}|^{2} \right)^{2} - 4 \gamma_{1}^{2} \omega^{2}}}{2 \gamma_{1}^{2} \omega}.
\end{equation}
For a (BA)-stacked half-crystal $\hat{G}^{\textrm{L,(BA)}}_{0,0}=\hat{G}^{\textrm{R,(AB)}}_{N+1,N+1}$ and $\hat{G}^{\textrm{R,(BA)}}_{N+1,N+1}=\hat{G}^{\textrm{L,(AB)}}_{0,0}$.

\newpage
\section{Full charge redistribution results: $\mathcal{R}$-$\mathcal{R}$ junctions}

To ensure overall charge neutrality better than $10^{-6}e$, we consider 20 layers either side of the physical interface for three of the the $\mathcal{R}$-$\mathcal{R}$ junctions, giving $N=40$. For the AB$|$AC junction we use $N=41$, to ensure mirror symmetry of the electric potential about layer $J$.

\begin{figure*}[h!]
\centering
\includegraphics[width=\textwidth]{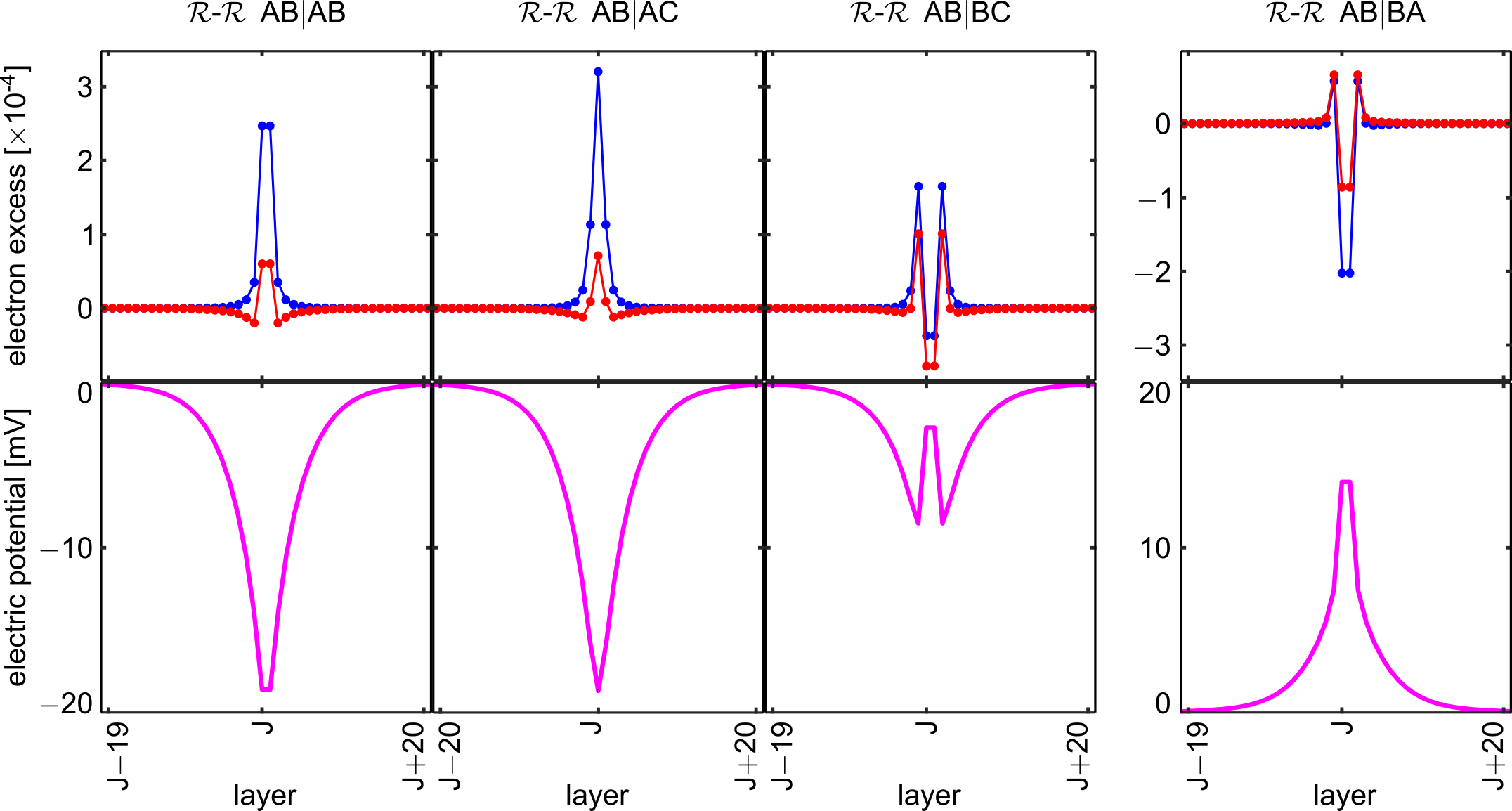}
\caption{The calculated electron excess and electric potential found in the vicinity of the $\mathcal{R}$-$\mathcal{R}$ junctions. Red (blue) are the number of excess electrons on each layer after (before) the self-consistent procedure, and electric potential is shown in magenta. All electron excess and electric potential results presented in this document use the same color scheme as used in this figure.}
\end{figure*}

\newpage
\section{Full charge redistribution results: $\mathcal{B}$-$\mathcal{R}$ junctions}

To ensure overall charge neutrality better than $10^{-6}e$ for the $\mathcal{B}$-$\mathcal{R}$ junctions, we consider 10 layers on the Bernal side of the physical interface and 20 layers on the rhombohedral side, giving $N=30$.

\begin{figure*}[h!]
\centering
\includegraphics[width=\textwidth]{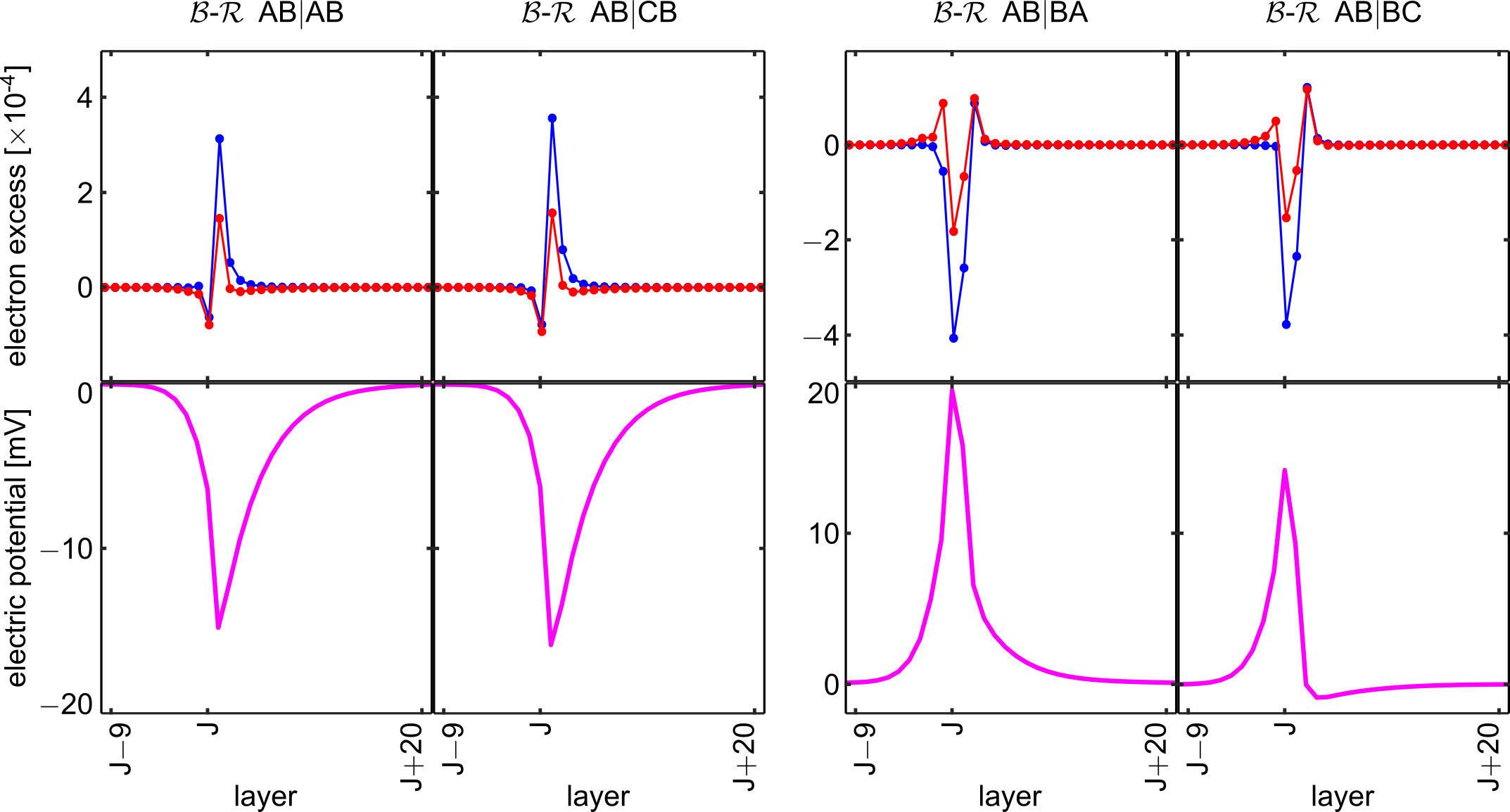}
\caption{The calculated electron excess and electric potential found in the vicinity of the $\mathcal{B}$-$\mathcal{R}$ junctions.}
\end{figure*}

\newpage
\section{Full charge redistribution results: $\mathcal{B}$-$\mathcal{B}$ junctions}

To ensure overall charge neutrality better than $10^{-6}e$, we consider 10 layers either side of the physical interface for three of the the $\mathcal{B}$-$\mathcal{B}$ junctions, giving $N=20$. For the AB$|$AC junction we take $N=21$, to ensure mirror symmetry of the electric potential about layer $J+1$.

\begin{figure*}[h!]
\centering
\includegraphics[width=\textwidth]{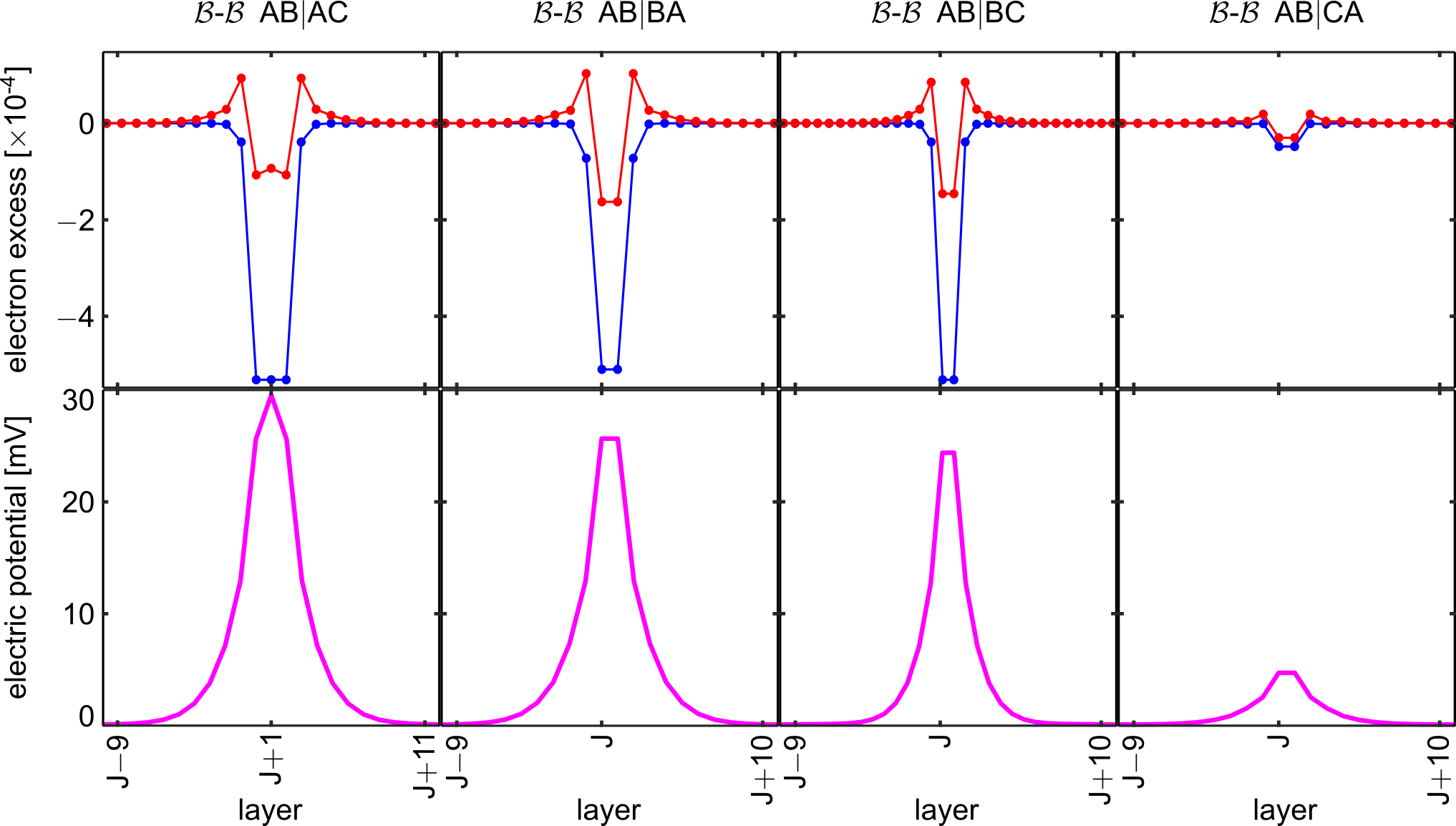}
\caption{The calculated electron excess and electric potential found in the vicinity of the $\mathcal{B}$-$\mathcal{B}$ junctions.}
\end{figure*}

\newpage
\section{Full DOS results: $\mathcal{R}$-$\mathcal{R}$ AB$|$AB junction}

\begin{figure*}[h!]
\centering
\includegraphics[width=\textwidth]{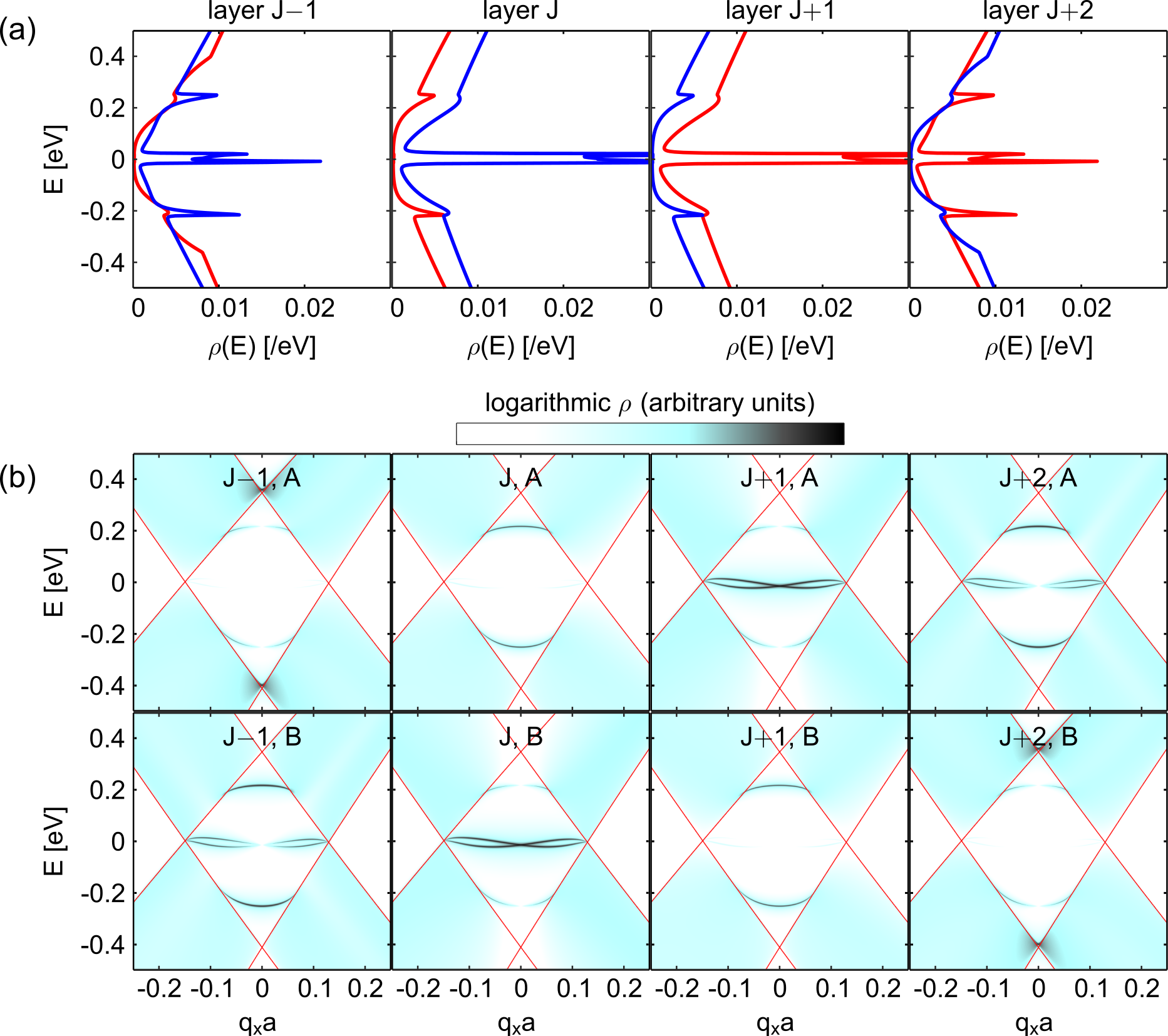}
\caption{a) The site-resolved density of states for the $\mathcal{R}$-$\mathcal{R}$ AB$|$AB junction. Results for sublattice $A$ ($B$) are shown in red (blue). b) The corresponding $\vect{q}$-resolved density of states. Darker areas correspond to higher LDOS. Here $\vect{q}=\vect{q}-\vect{K}$ is a wave vector measured from the centre of the valley $K$, and $q_{y}=0$ in these results. All $\vect{q}$-resolved DOS results presented in this document use the same scale and colour scheme as used in this figure.}
\end{figure*}

\newpage
\section{Full DOS results: $\mathcal{R}$-$\mathcal{R}$ AB$|$AC junction}

\begin{figure*}[h!]
\centering
\includegraphics[width=\textwidth]{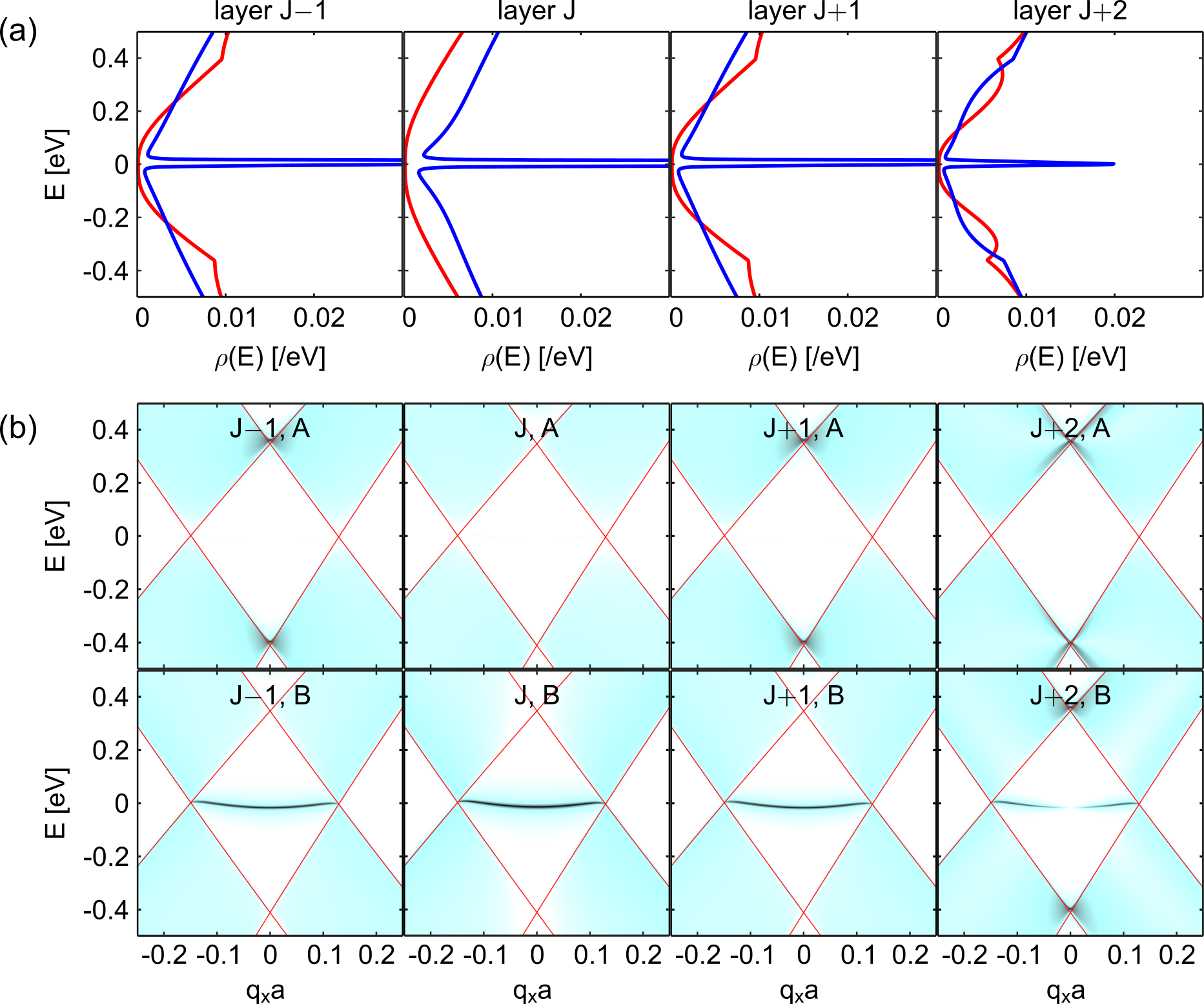}
\caption{The a) site-resolved and b) $\vect{q}$-resolved density of states for the $\mathcal{R}$-$\mathcal{R}$ AB$|$AC junction.}
\end{figure*}

\newpage
\section{Full DOS results: $\mathcal{R}$-$\mathcal{R}$ AB$|$BA junction}

\begin{figure*}[h!]
\centering
\includegraphics[width=\textwidth]{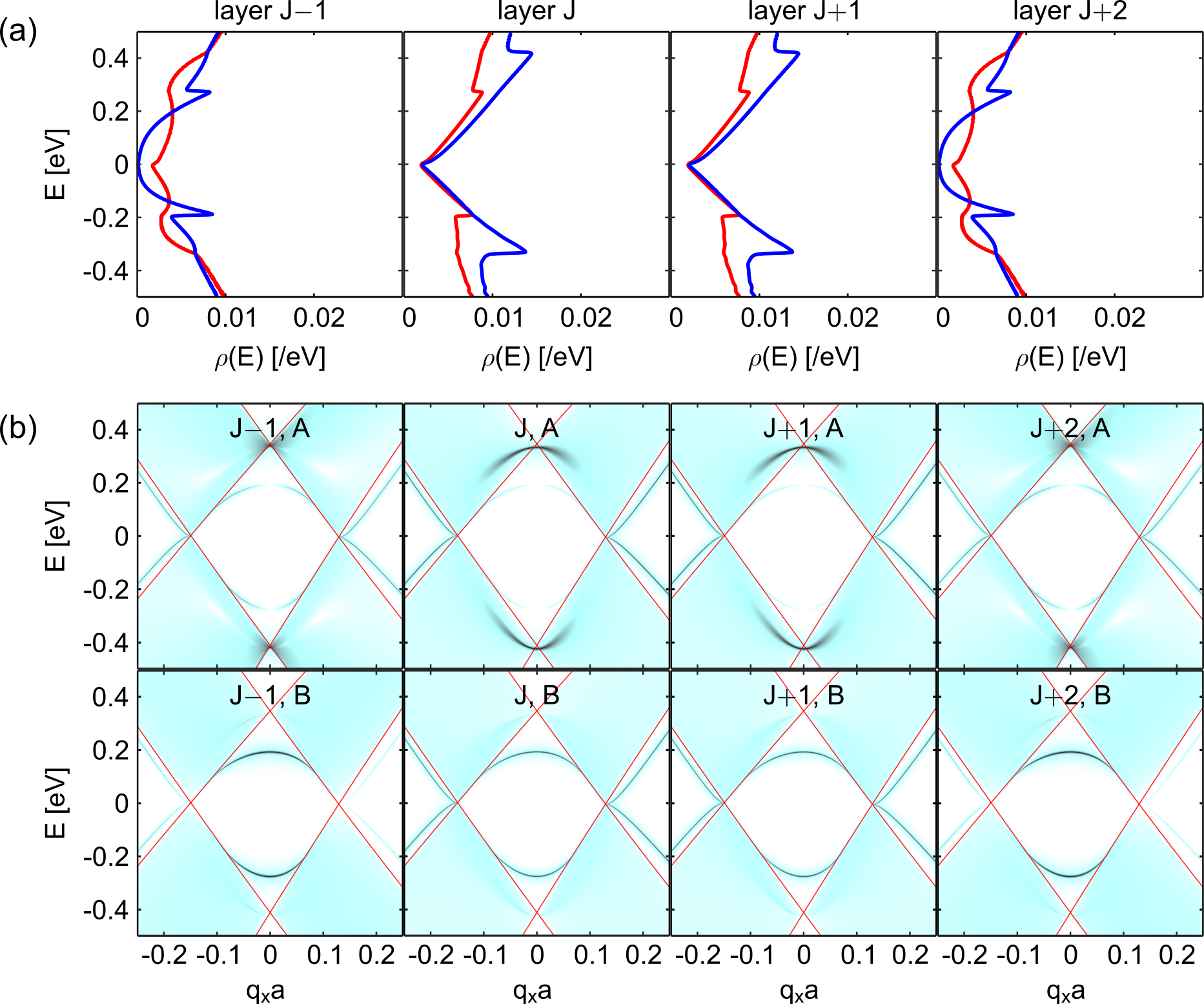}
\caption{The a) site-resolved and b) $\vect{q}$-resolved density of states for the $\mathcal{R}$-$\mathcal{R}$ AB$|$BA junction.}
\end{figure*}

\newpage
\section{Full DOS results: $\mathcal{R}$-$\mathcal{R}$ AB$|$BC junction}

\begin{figure*}[h!]
\centering
\includegraphics[width=\textwidth]{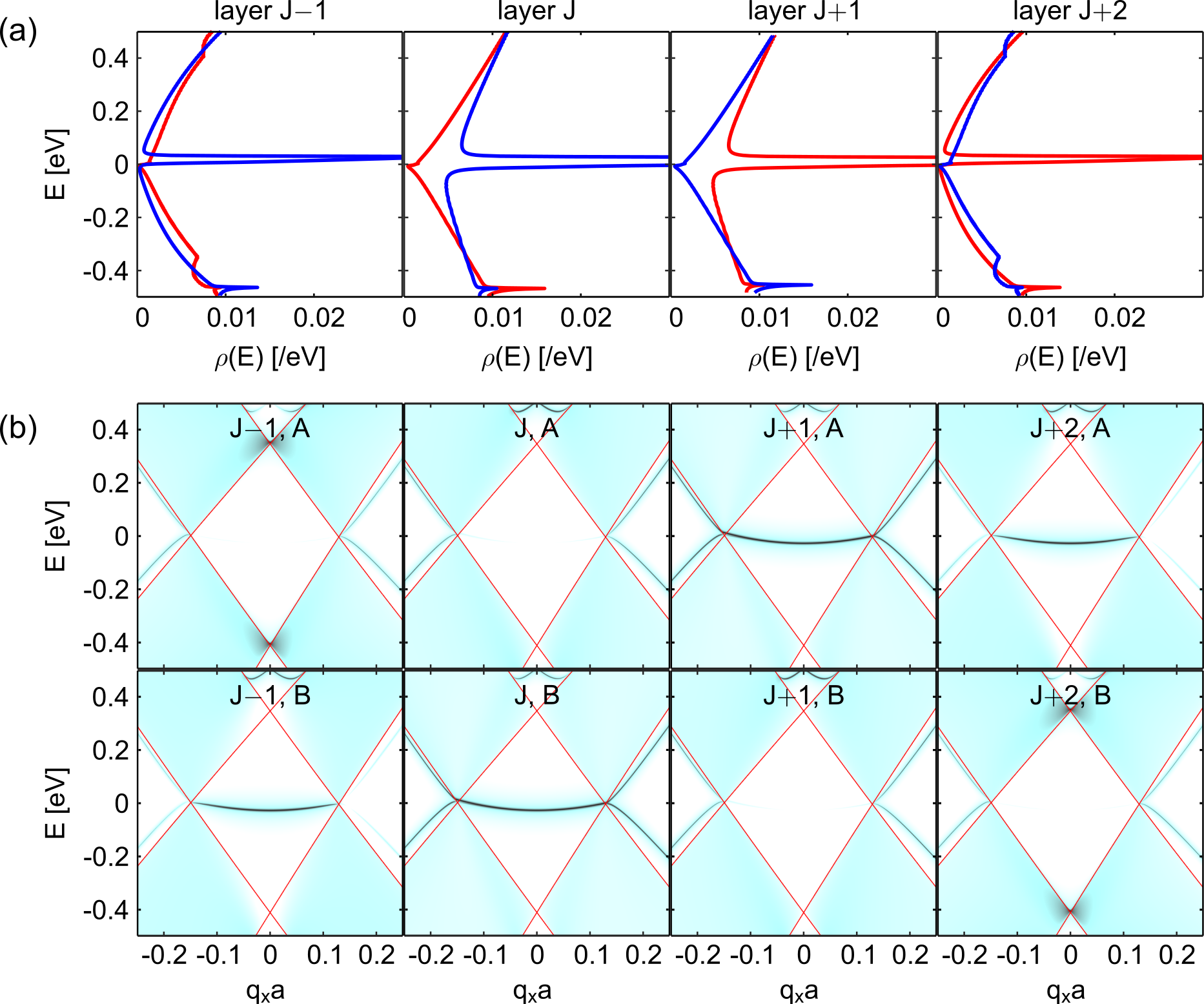}
\caption{The a) site-resolved and b) $\vect{q}$-resolved density of states for the $\mathcal{R}$-$\mathcal{R}$ AB$|$BC junction.}
\end{figure*}

\newpage
\section{Full DOS results: $\mathcal{B}$-$\mathcal{R}$ AB$|$AB junction}

\begin{figure*}[h!]
\centering
\includegraphics[width=\textwidth]{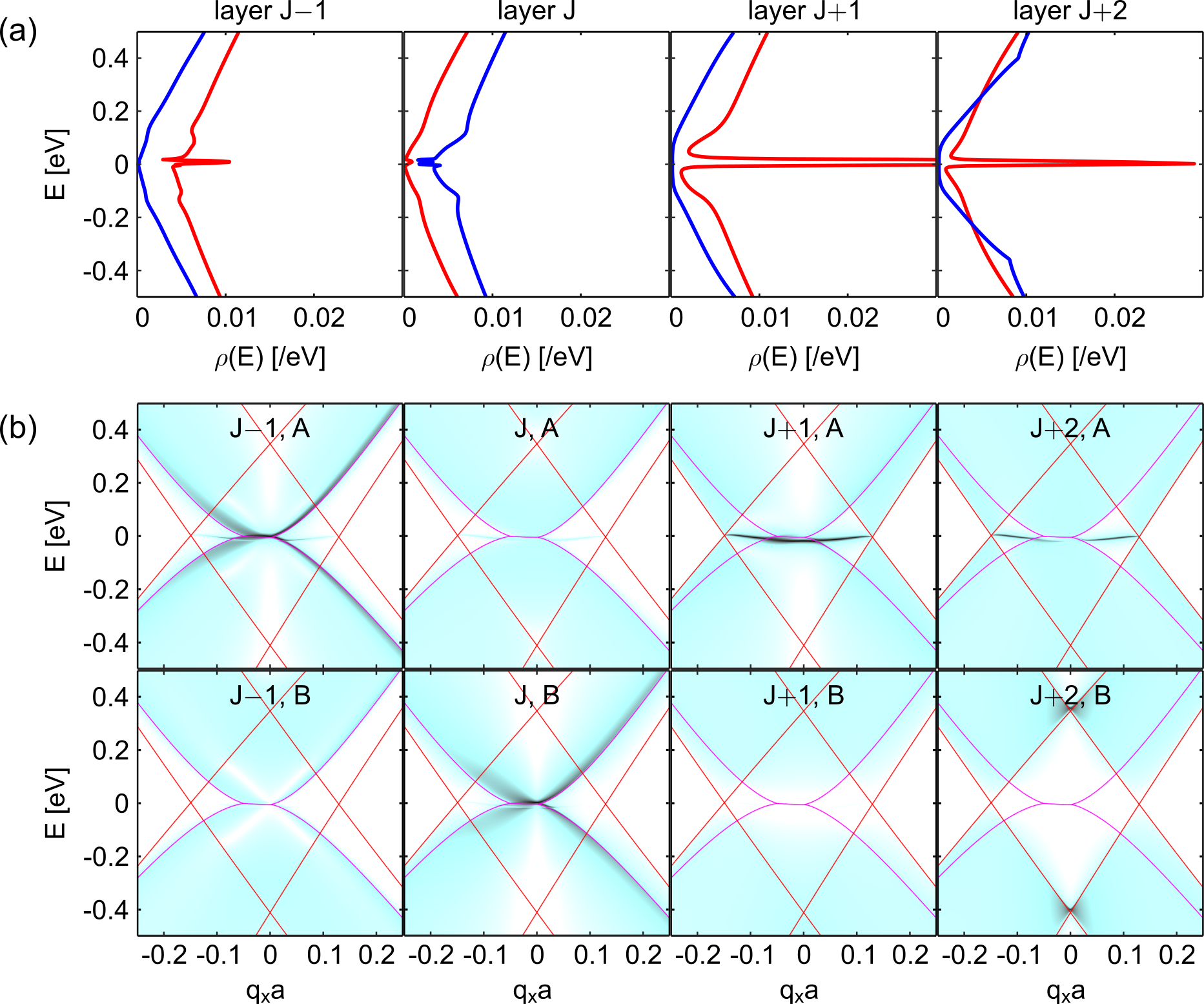}
\caption{The a) site-resolved and b) $\vect{q}$-resolved density of states for the $\mathcal{B}$-$\mathcal{R}$ AB$|$AB junction.}
\end{figure*}

\newpage
\section{Full DOS results: $\mathcal{B}$-$\mathcal{R}$ AB$|$BA junction}

\begin{figure*}[h!]
\centering
\includegraphics[width=\textwidth]{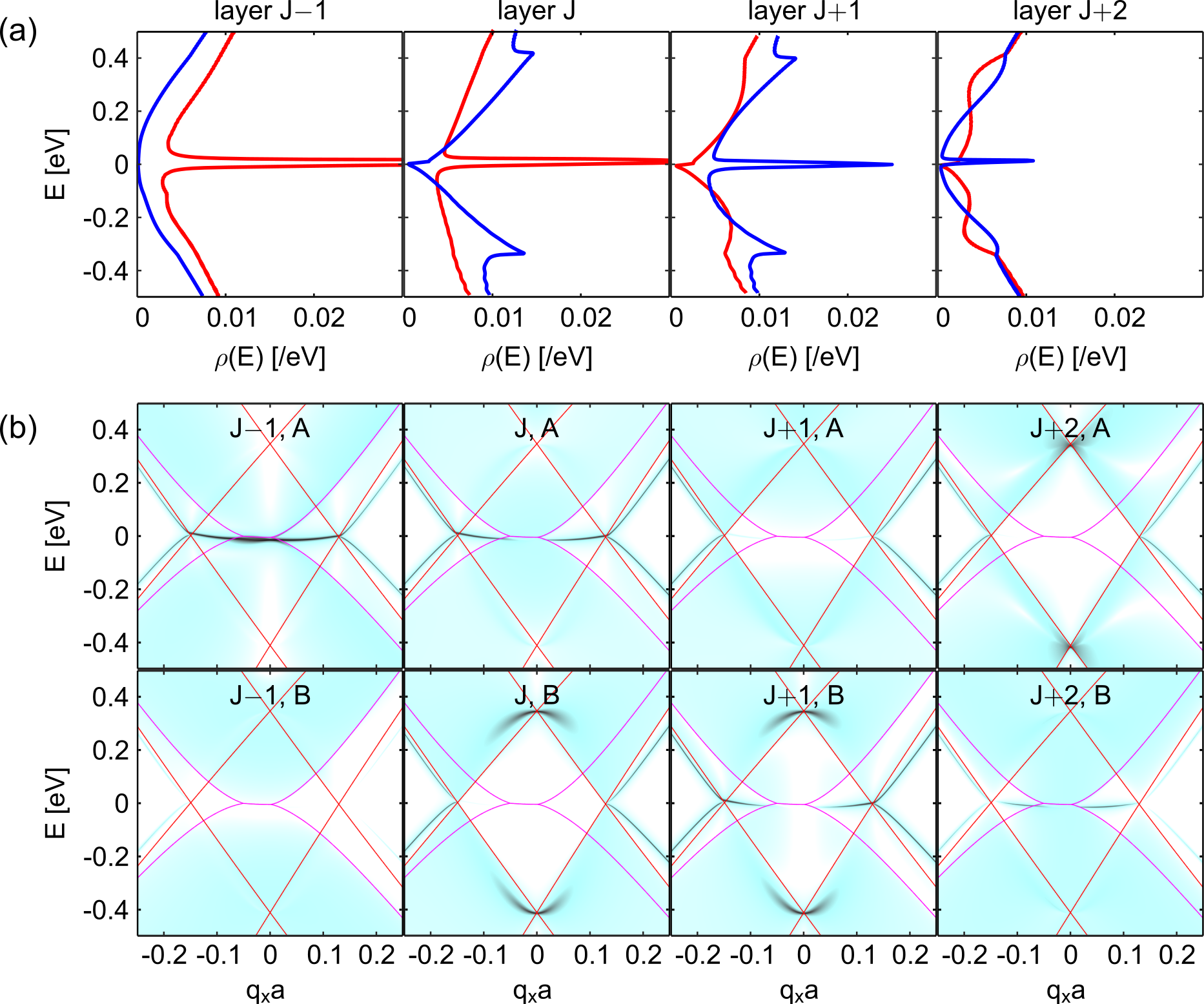}
\caption{The a) site-resolved and b) $\vect{q}$-resolved density of states for the $\mathcal{B}$-$\mathcal{R}$ AB$|$BA junction.}
\end{figure*}

\newpage
\section{Full DOS results: $\mathcal{B}$-$\mathcal{R}$ AB$|$BC junction}

\begin{figure*}[h!]
\centering
\includegraphics[width=\textwidth]{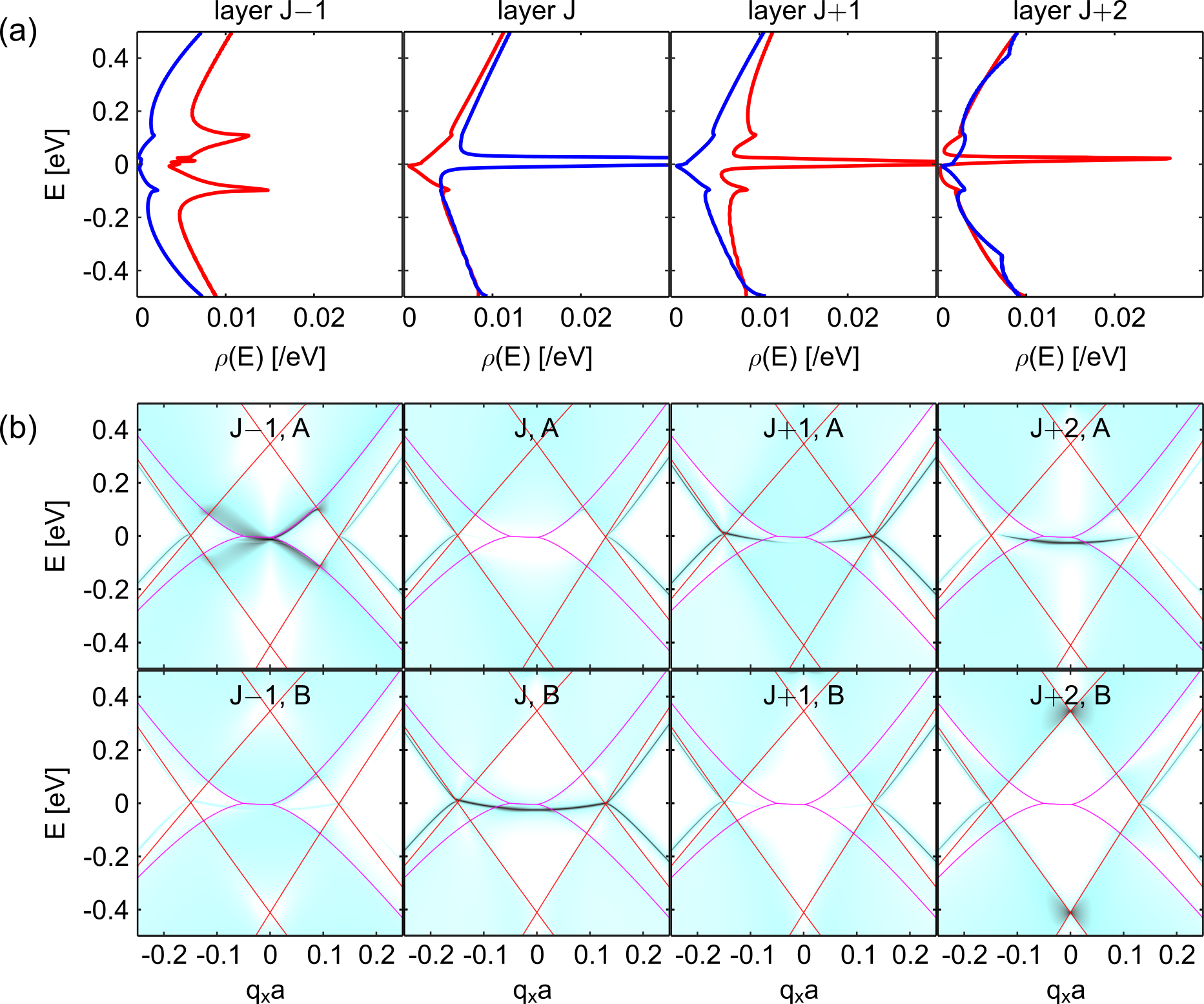}
\caption{The a) site-resolved and b) $\vect{q}$-resolved density of states for the $\mathcal{B}$-$\mathcal{R}$ AB$|$BC junction.}
\end{figure*}

\newpage
\section{Full DOS results: $\mathcal{B}$-$\mathcal{R}$ AB$|$CB junction}

\begin{figure*}[h!]
\centering
\includegraphics[width=\textwidth]{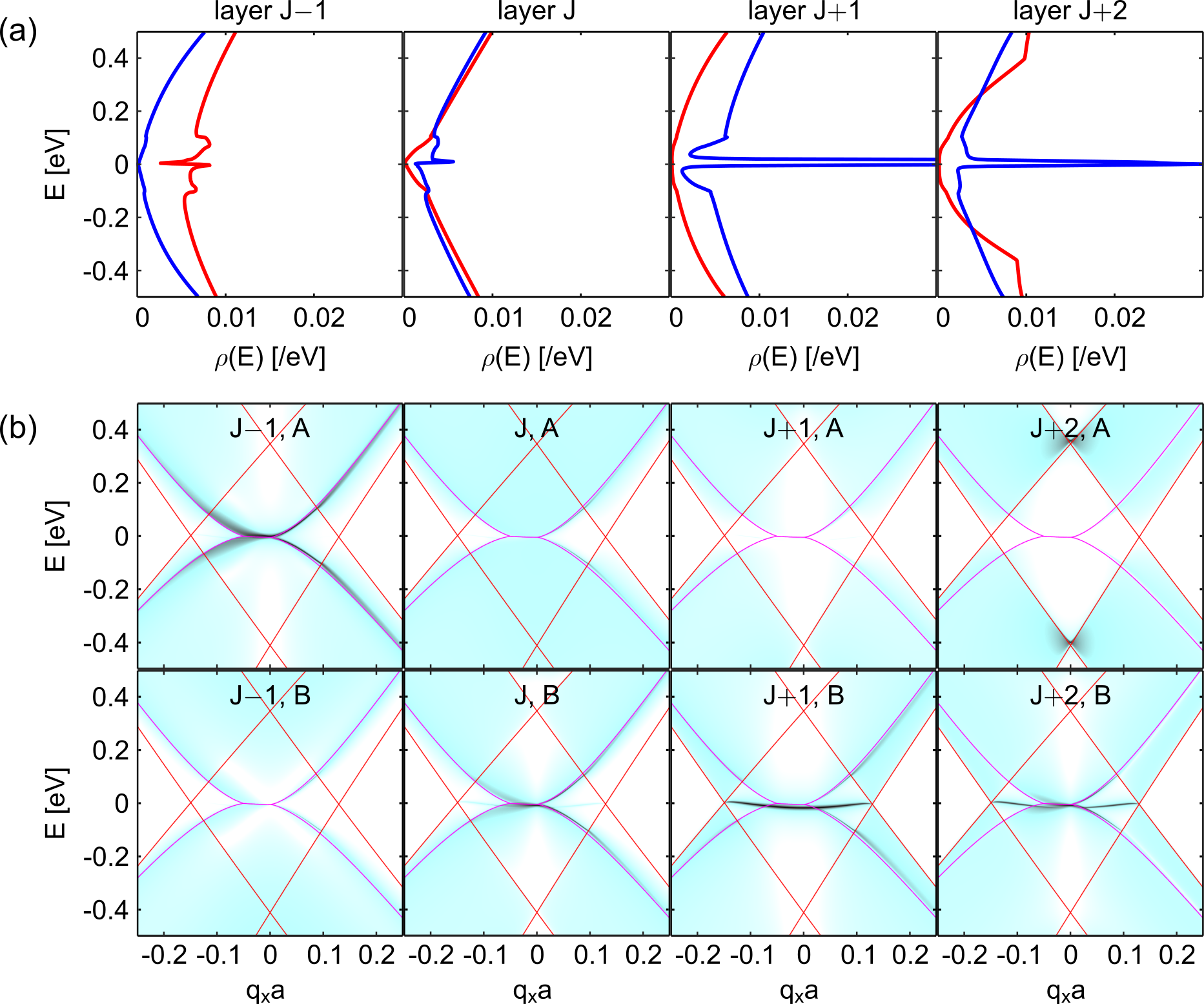}
\caption{The a) site-resolved and b) $\vect{q}$-resolved density of states for the $\mathcal{B}$-$\mathcal{R}$ AB$|$CB junction.}
\end{figure*}

\newpage
\section{Full DOS results: $\mathcal{B}$-$\mathcal{B}$ AB$|$AC junction}

\begin{figure*}[h!]
\centering
\includegraphics[width=\textwidth]{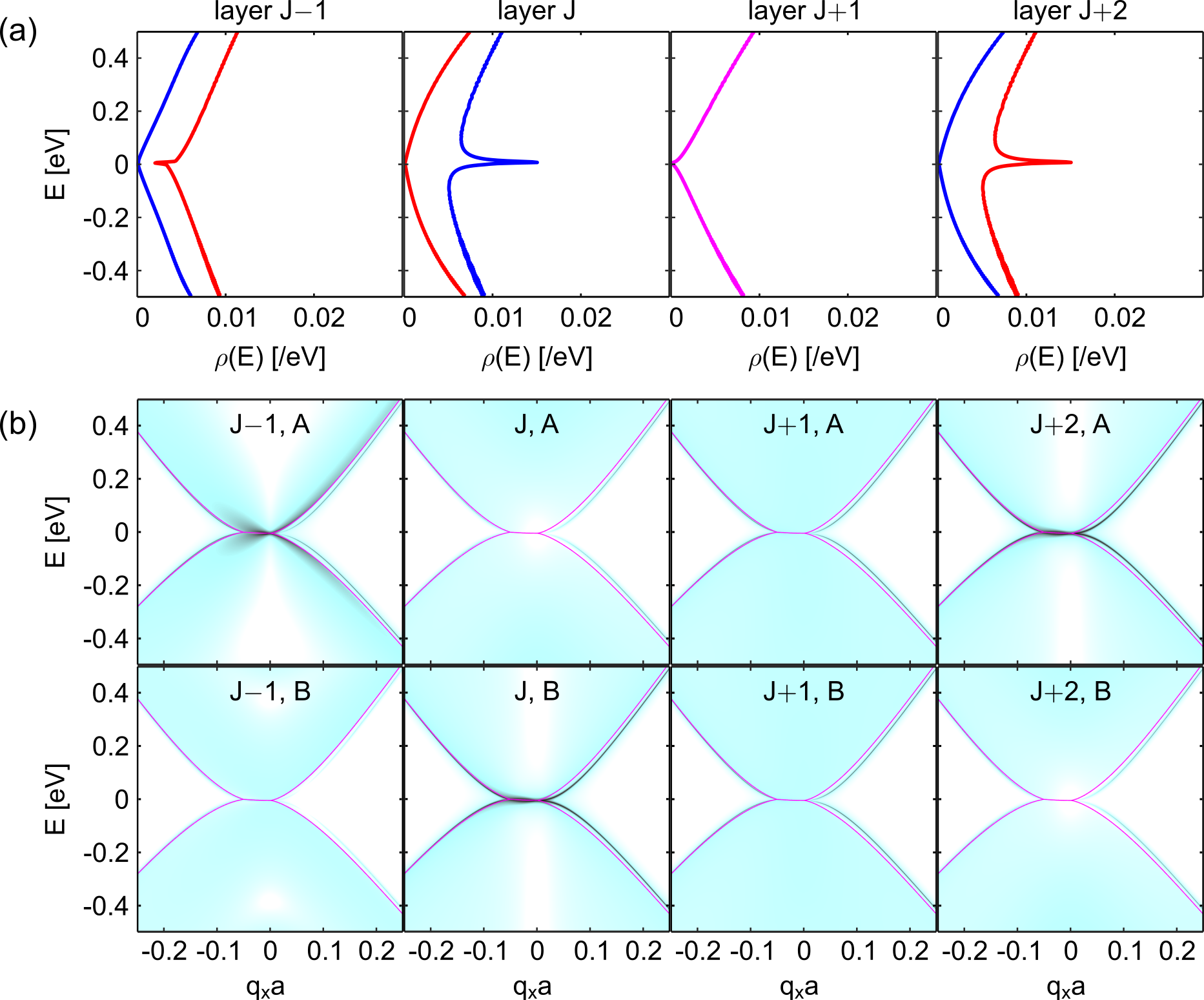}
\caption{The a) site-resolved and b) $\vect{q}$-resolved density of states for the $\mathcal{B}$-$\mathcal{B}$ AB$|$AC junction. Site-resolved results are shown in magenta when the sublattices are equal.}
\end{figure*}

\newpage
\section{Full DOS results: $\mathcal{B}$-$\mathcal{B}$ AB$|$BA junction}

\begin{figure*}[h!]
\centering
\includegraphics[width=\textwidth]{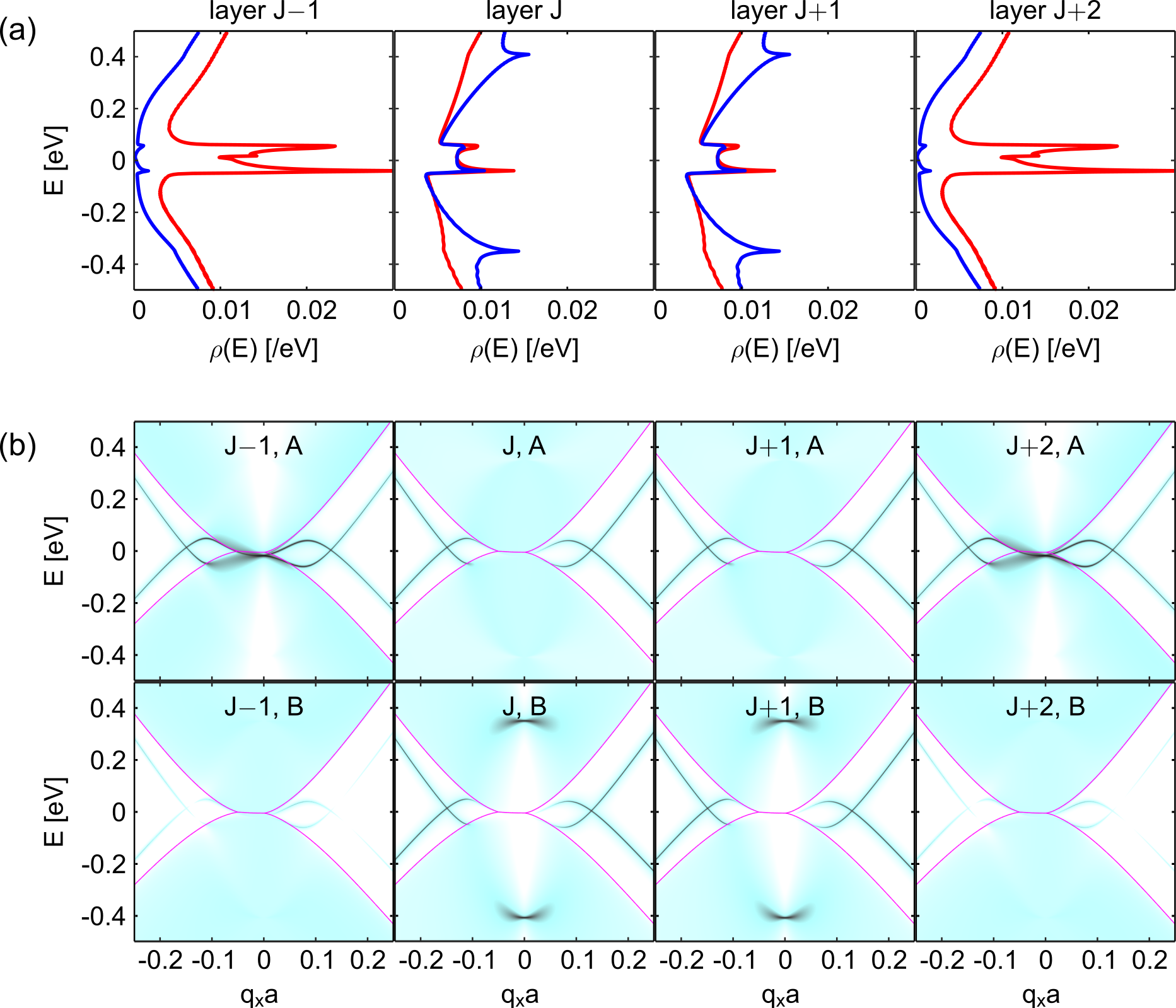}
\caption{The a) site-resolved and b) $\vect{q}$-resolved density of states for the $\mathcal{B}$-$\mathcal{B}$ AB$|$BA junction.}
\end{figure*}

\newpage
\section{Full DOS results: $\mathcal{B}$-$\mathcal{B}$ AB$|$BC junction}

\begin{figure*}[h!]
\centering
\includegraphics[width=\textwidth]{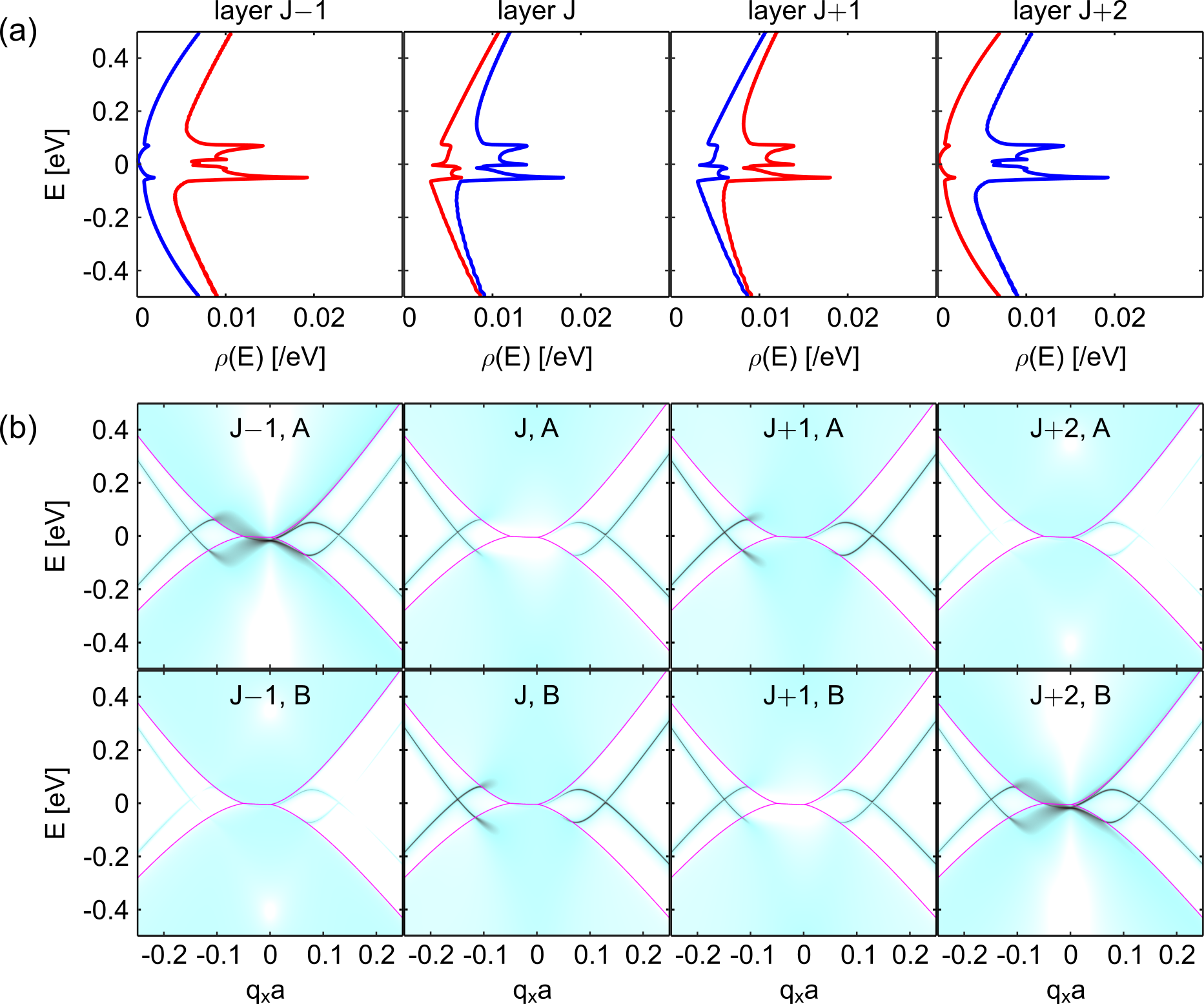}
\caption{The a) site-resolved and b) $\vect{q}$-resolved density of states for the $\mathcal{B}$-$\mathcal{B}$ AB$|$BC junction.}
\end{figure*}

\newpage
\section{Full DOS results: $\mathcal{B}$-$\mathcal{B}$ AB$|$CA junction}

\begin{figure*}[h!]
\centering
\includegraphics[width=\textwidth]{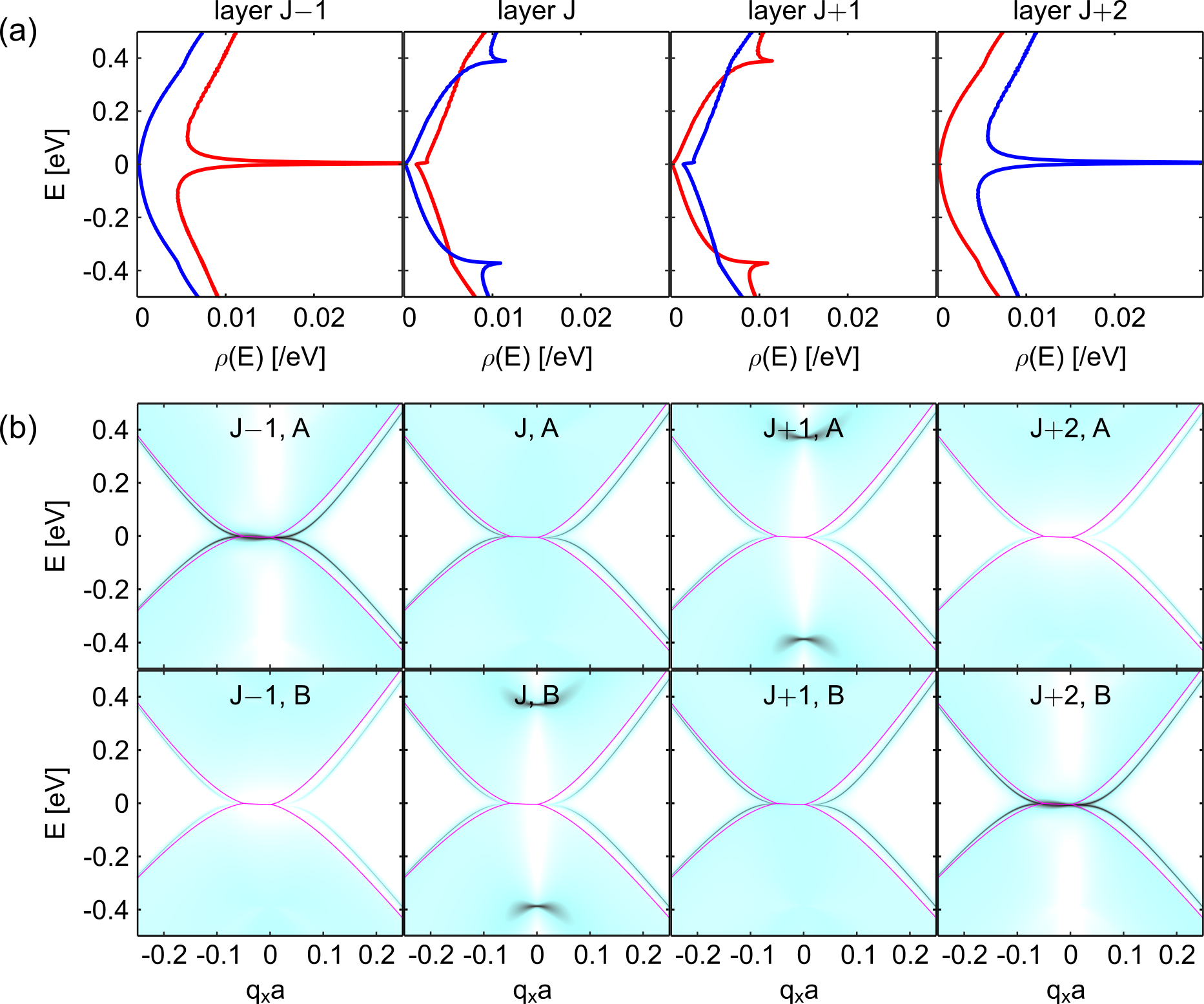}
\caption{The a) site-resolved and b) $\vect{q}$-resolved density of states for the $\mathcal{B}$-$\mathcal{B}$ AB$|$CA junction.}
\end{figure*}

\newpage
\section{Analytical dispersion relations}

The dispersion relation for the dispersive states at particular junctions is determined from the poles of the Green's function in the minimal model (considering $\gamma_{0}$ and $\gamma_{1}$ only). Analytical expressions for each of the junctions that exhibit such states are given below, and are plotted in Supplemental Fig.\@ 16.

$\mathcal{B}$-$\mathcal{B}$ AB$|$AC:
\begin{equation}
|E| = \frac{\gamma_{1}^{2}+2\gamma_{0}^{2}|f_{\vect{q}}|^{2}-\gamma_{1}\sqrt{\gamma_{1}^{2}+4\gamma_{0}^{2}|f_{\vect{q}}|^{2}}}{2 \gamma_{0}|f_{\vect{q}}|}.
\end{equation}

$\mathcal{B}$-$\mathcal{B}$ AB$|$BA:
\begin{equation}
\begin{split}
\gamma_{0}^{2}|f_{\vect{q}}|^{2} &= \frac{\left(\gamma_{1} + \sqrt{\gamma_{1}^{2} + 8\gamma_{1}|E| + 4|E|^{2}} \right)\left(\gamma_{1} + |E|\right)}{2}, \\
\gamma_{0}^{2}|f_{\vect{q}}|^{2} &= \frac{\left(\gamma_{1} + \sqrt{\gamma_{1}^{2} - 8\gamma_{1}|E| + 4|E|^{2}} \right)\left(\gamma_{1} - |E|\right)}{2}, \\
\gamma_{0}^{2}|f_{\vect{q}}|^{2} &= \frac{\left(\gamma_{1} - \sqrt{\gamma_{1}^{2} - 8\gamma_{1}|E| + 4|E|^{2}} \right)\left(\gamma_{1} - |E|\right)}{2},
\end{split}
\end{equation}

$\mathcal{B}$-$\mathcal{B}$ AB$|$BC:
\begin{equation}
\begin{split}
|E| &= \frac{(\gamma_{1}-\gamma_{0}|f_{\vect{q}}|)\sqrt{2\gamma_{0}|f_{\vect{q}}|(2\gamma_{0}|f_{\vect{q}}|-\gamma_{1})}}{2 \gamma_{0}|f_{\vect{q}}|}, \ |f_{\vect{q}}| > \frac{\gamma_{1}}{\sqrt{3}\gamma_{0}}, \\
|E| &= \frac{(\gamma_{0}|f_{\vect{q}}|-\gamma_{1})\sqrt{2\gamma_{0}|f_{\vect{q}}|(2\gamma_{0}|f_{\vect{q}}|-\gamma_{1})}}{2 \gamma_{0}|f_{\vect{q}}|}.
\end{split}
\end{equation}

$\mathcal{B}$-$\mathcal{B}$ AB$|$CA:
\begin{equation}
\gamma_{0}^{2}|f_{\vect{q}}|^{2} = \frac{\left(|E|+\sqrt{4\gamma_{1}|E|+|E|^{2}} \right)(\gamma_{1}+|E|)}{2}.
\end{equation}

$\mathcal{R}$-$\mathcal{R}$ AB$|$AB:
\begin{equation}
\gamma_{0}^{2}|f_{\vect{q}}|^{2} = \frac{|E| \left( |E|^{2} + |E|\gamma_{1} - \gamma_{1}^{2} \right)}{|E|-\gamma_{1}}, \ |E| > \frac{\gamma_{1}}{2}.
\end{equation}

$\mathcal{R}$-$\mathcal{R}$ AB$|$BA:
\begin{equation}
\begin{split}
\gamma_{0}^{2}|f_{\vect{q}}|^{2} &= \gamma_{1} \left(\gamma_{1} + |E|\right) + \sqrt{|E| \left(\gamma_{1} + |E|\right)^{3}}, \\
\gamma_{0}^{2}|f_{\vect{q}}|^{2} &= \gamma_{1} \left(\gamma_{1} + |E|\right) - \sqrt{|E| \left(\gamma_{1} + |E|\right)^{3}}, \ |E| > \frac{\gamma_{1}}{3},
\end{split}
\end{equation}

$\mathcal{R}$-$\mathcal{R}$ AB$|$BC:
\begin{equation}
|E| = \frac{\sqrt{2} \left(\gamma_{0}|f_{\vect{q}}|-\gamma_{1}\right)^{\frac{3}{2}}}{\left(2\gamma_{0}|f_{\vect{q}}|-\gamma_{1}\right)^{\frac{1}{2}}}.
\end{equation}

Dispersive states are also observed in the $\mathcal{B}$-$\mathcal{R}$ AB$|$BA, $\mathcal{B}$-$\mathcal{R}$ AB$|$BC, and $\mathcal{B}$-$\mathcal{R}$ AB$|$CB crystals, but no simple analytical expressions for their dispersion relations exist.

\begin{figure*}[h!]
\centering
\includegraphics[width=\textwidth]{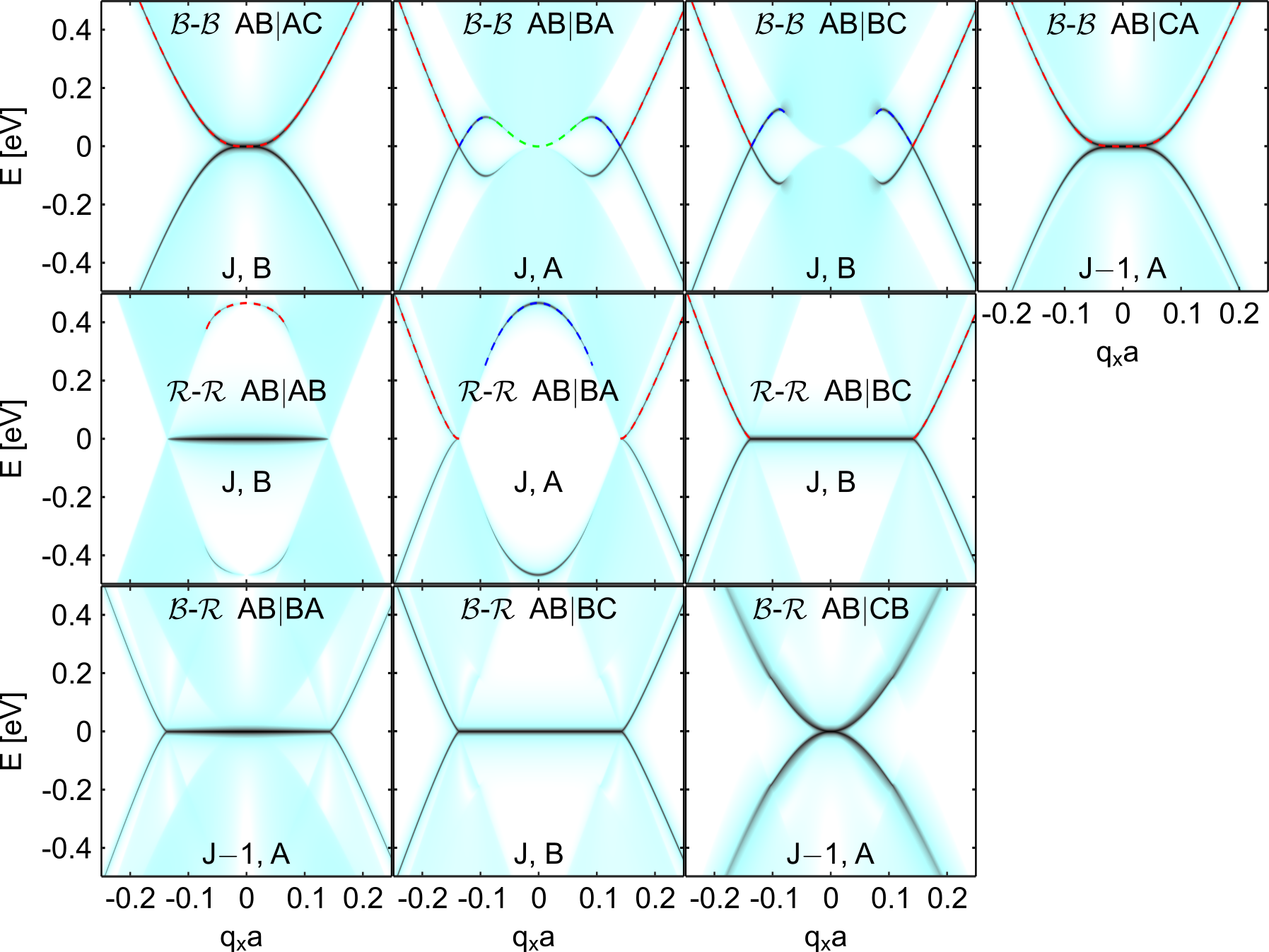}
\caption{The $\vect{q}$-resolved density of states on atoms that most strongly host dispersive states. A smaller energy range than the previous LDOS figures is used here, as we only present results for low-energy junction states in the range $|E|\lesssim\gamma_{1}$. The dispersion relations are shown with dashed lines; these are symmetric in energy about $E=0$ but only the dispersion for $E>0$ is shown. For states described by more than one expression, red, blue and green are used in the order that the expressions are presented in the text.}
\end{figure*}

\newpage
\section{Analytical zero-energy states}

An analytical expression for the number of zero-energy states (per spin per valley) $n_{j}^{\mu}$ on sublattice $\mu$ in layer $j$ was obtained from a low-energy expansion of the Green's function in the minimal model. These are given below for all junctions that exhibit such states, and plotted in Fig.\@ 6 of the main text.

$\mathcal{R}$-$\mathcal{R}$ AB$|$AB:
\begin{equation}
n^{A}_{J+m+1}(\vect{q})=n^{B}_{J-m}(\vect{q}) = \frac{\left(\gamma_{0}^{2}|f_{\vect{q}}|^{2}\right)^{m} \left(\gamma_{1}^{2}-\gamma_{0}^{2}|f_{\vect{q}}|^{2}\right)}{\gamma_{1}^{2(m+1)}} \Theta \left(\gamma_{1}^{2}-\gamma_{0}^{2}|f_{\vect{q}}|^{2}\right), \ \ m \geq 0.
\end{equation}

$\mathcal{R}$-$\mathcal{R}$ AB$|$AC:
\begin{equation}
\begin{split}
n^{B}_{J}(\vect{q}) &= \frac{2\left(\gamma_{1}^{2}-\gamma_{0}^{2}|f_{\vect{q}}|^{2}\right)}{2\gamma_{1}^{2}-\gamma_{0}^{2}|f_{\vect{q}}|^{2}} \Theta \left(\gamma_{1}^{2}-\gamma_{0}^{2}|f_{\vect{q}}|^{2}\right), \\
n^{B}_{J \pm m}(\vect{q}) &= \frac{\left(\gamma_{0}^{2}|f_{\vect{q}}|^{2}\right)^{m-1} \left(\gamma_{1}^{2}-\gamma_{0}^{2}|f_{\vect{q}}|^{2}\right)}{\gamma_{1}^{2(m-1)} \left(2\gamma_{1}^{2}-\gamma_{0}^{2}|f_{\vect{q}}|^{2}\right)} \Theta \left(\gamma_{1}^{2}-\gamma_{0}^{2}|f_{\vect{q}}|^{2}\right), \ \ m \geq 1.
\end{split}
\end{equation}

$\mathcal{R}$-$\mathcal{R}$ AB$|$BC:
\begin{equation}
\begin{split}
n^{A}_{J+1}(\vect{q})=n^{B}_{J}(\vect{q}) &= \frac{1}{2} \Theta \left(\gamma_{1}^{2}-\gamma_{0}^{2}|f_{\vect{q}}|^{2}\right), \\
n^{A}_{J+m+1}(\vect{q})=n^{B}_{J-m}(\vect{q}) &= \frac{\left(\gamma_{0}^{2}|f_{\vect{q}}|^{2}\right)^{m-1} \left(\gamma_{1}^{2}-\gamma_{0}^{2}|f_{\vect{q}}|^{2}\right)}{2\gamma_{1}^{2m}} \Theta \left(\gamma_{1}^{2}-\gamma_{0}^{2}|f_{\vect{q}}|^{2}\right), \ \ m \geq1.
\end{split}
\end{equation}

$\mathcal{B}$-$\mathcal{R}$ AB$|$AB:
\begin{equation}
n^{A}_{J+m+1}(\vect{q}) = \frac{\left(\gamma_{0}^{2}|f_{\vect{q}}|^{2}\right)^{m} \left(\gamma_{1}^{2}-\gamma_{0}^{2}|f_{\vect{q}}|^{2}\right)}{\gamma_{1}^{2(m+1)}} \Theta \left(\gamma_{1}^{2}-\gamma_{0}^{2}|f_{\vect{q}}|^{2}\right), \ \ m \geq 0.
\end{equation}

$\mathcal{B}$-$\mathcal{R}$ AB$|$BA:
\begin{equation}
\begin{split}
n^{A}_{J-1}(\vect{q}) &= \frac{\gamma_{1}^{2}}{\gamma_{1}^{2}+2\gamma_{0}^{2}|f_{\vect{q}}|^{2}} \Theta \left(\gamma_{1}^{2}-\gamma_{0}^{2}|f_{\vect{q}}|^{2}\right), \\
n^{A}_{J}(\vect{q}) &= \frac{\gamma_{0}^{2}|f_{\vect{q}}|^{2}}{\gamma_{1}^{2}+2\gamma_{0}^{2}|f_{\vect{q}}|^{2}} \Theta \left(\gamma_{1}^{2}-\gamma_{0}^{2}|f_{\vect{q}}|^{2}\right), \\
n^{B}_{J+1}(\vect{q}) &= \frac{\gamma_{0}^{4}|f_{\vect{q}}|^{4}}{\gamma_{1}^{2}\left(\gamma_{1}^{2}+2\gamma_{0}^{2}|f_{\vect{q}}|^{2}\right)} \Theta \left(\gamma_{1}^{2}-\gamma_{0}^{2}|f_{\vect{q}}|^{2}\right), \\
n^{B}_{J+m}(\vect{q}) &= \frac{\left(\gamma_{0}^{2}|f_{\vect{q}}|^{2}\right)^{2(m-1)}\left(\gamma_{1}^{2}-\gamma_{0}^{2}|f_{\vect{q}}|^{2}\right)^{2}}{\gamma_{1}^{2m}\left(\gamma_{1}^{2}+2\gamma_{0}^{2}|f_{\vect{q}}|^{2}\right)} \Theta \left(\gamma_{1}^{2}-\gamma_{0}^{2}|f_{\vect{q}}|^{2}\right), \ \ m \geq 2.
\end{split}
\end{equation}

$\mathcal{B}$-$\mathcal{R}$ AB$|$BC:
\begin{equation}
\begin{split}
n^{B}_{J}(\vect{q}) &= \frac{1}{2} \Theta \left(\gamma_{1}^{2}-\gamma_{0}^{2}|f_{\vect{q}}|^{2}\right), \\
n^{A}_{J+1}(\vect{q}) &= \frac{\gamma_{0}^{2}|f_{\vect{q}}|^{2}}{2\gamma_{1}^{2}} \Theta \left(\gamma_{1}^{2}-\gamma_{0}^{2}|f_{\vect{q}}|^{2}\right), \\
n^{A}_{J+m}(\vect{q}) &= \frac{\left(\gamma_{0}^{2}|f_{\vect{q}}|^{2}\right)^{m-2}\left(\gamma_{1}^{2}-\gamma_{0}^{2}|f_{\vect{q}}|^{2}\right)^{2}}{2\gamma_{1}^{2m}} \Theta \left(\gamma_{1}^{2}-\gamma_{0}^{2}|f_{\vect{q}}|^{2}\right), \ \ m \geq 2.
\end{split}
\end{equation}

$\mathcal{B}$-$\mathcal{R}$ AB$|$CB:
\begin{equation}
n^{B}_{J+m+1}(\vect{q}) = \frac{\left(\gamma_{0}^{2}|f_{\vect{q}}|^{2}\right)^{m} \left(\gamma_{1}^{2}-\gamma_{0}^{2}|f_{\vect{q}}|^{2}\right)}{\gamma_{1}^{2(m+1)}} \Theta \left(\gamma_{1}^{2}-\gamma_{0}^{2}|f_{\vect{q}}|^{2}\right), \ \ m \geq 0.
\end{equation}

\newpage
\section{Effect of size of rhombohedral regions on junction states}

The $\mathcal{B}$-$\mathcal{R}$ AB$|$CB junction (results shown in Supplemental Fig.\@ 11) provides an opportunity to study the effect of the size of a rhombohedral region on the associated junction states, as it contains both a semi-infinite rhombohedral region and a rhombohedral trilayer. We investigate this by considering a finite rhombohedral stack of size $N$ embedded between a Bernal stacked half-crystal and an inverted rhombohedral stacked half-crystal, such that the $\mathcal{B}$-$\mathcal{R}$ AB$|$CB crystal is recovered when $N=3$. This crystal can be thought of as a $\mathcal{B}$-$\mathcal{R}$-$\mathcal{R}$ crystal, possessing an AB$|$AB junction on the left side of the finite rhombohedral region, and an AB$|$CB (or AB$|$AC)  junction on the right---see Supplemental Fig.\@ \ref{fig:brrcrystal}a.

\begin{figure}[t]
\centering
\includegraphics[width=0.6\columnwidth]{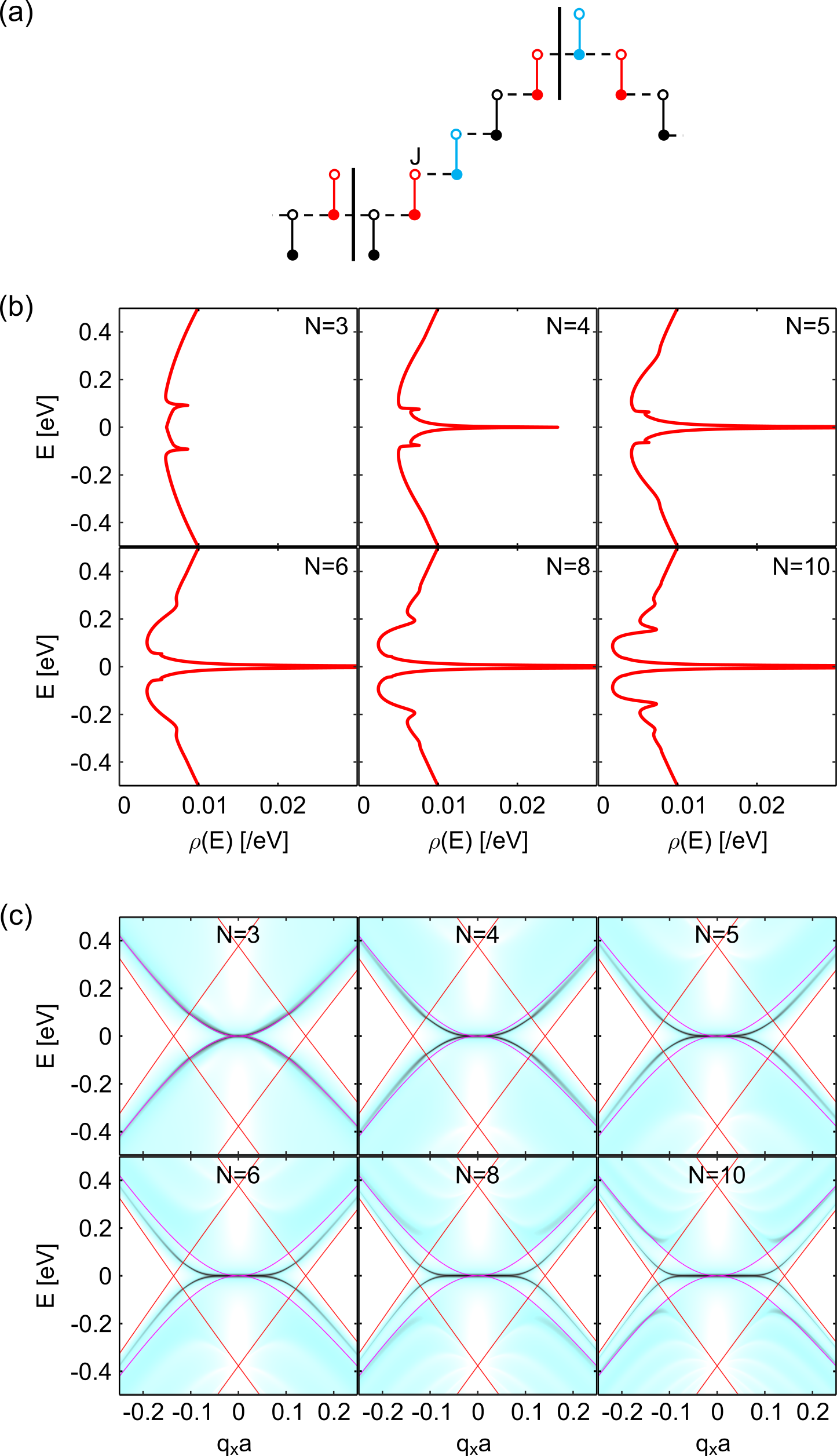}
\caption{\label{fig:brrcrystal} a) Schematic diagram of the $\mathcal{B}$-$\mathcal{R}$-$\mathcal{R}$ crystal for $N=6$. Layer $J$ is marked such that labeling remains consistent with the $\mathcal{B}$-$\mathcal{R}$ AB$|$CB junction for $N=3$. b) The site-resolved density of states on sublattice $A$ in layer $J-1$ for a selection of $N$. c) The corresponding $\vect{q}$-resolved density of states for $q_{y}=0$.}
\end{figure}

Results for the LDOS of this crystal for a selection of $N$ are given in Supplemental Fig.\@ \ref{fig:brrcrystal}b-c; these were calculated in the minimal model to avoid complications arising from charge redistribution at both junctions. We see that for $N \geq 4$, a spike emerges in the site-resolved LDOS of sublattice $A$ in layer $J-1$ at $E=0$. For $N \rightarrow \infty$, layer $J-1$ becomes effectively equivalent to layer $J+1$ in the $\mathcal{B}$-$\mathcal{R}$ AB$|$AB crystal---see Supplemental Fig.\@ 8. Analysis of the junction state on layer $J-1$ in the vicinity of $K$ uncovers an unusual low-energy dispersion relation of the form $|E| \propto  |\vect{q}|^{N-1}$, with corresponding site-resolved DOS of the form $\rho(E) \propto |E|^{\tfrac{2}{N-1}-1}$. From this it follows that, for $N \geq 4$, $\rho(E)$ diverges at $E=0$ and a spike is observed in the results. No spike in the site-resolved DOS is observed for $N=3$; this behavior arises as a consequence of the parabolic dispersion generating a constant $\rho(E)$, and not the absence of a rhombohedral edge state.